# Spatio-temporal optical vortices: principles of description and basic properties

A. BEKSHAEV

*Physics Research Institute, I.I. Mechnikov National University, Dvorianska 2, 65082, Odesa, Ukraine*
*\*bekshaev@onu.edu.ua*

**ABSTRACT**

This compilation represents a summary of the main physical foundations underlying the structure and properties of spatio-temporal optical vortices (STOV). The general approach to the STOV description and characterization is based on the model of scalar paraxial Gaussian wave packet. On this ground, the STOV structures of arbitrary orders are considered as superpositions of spatio-temporal Hermite-Gaussian modes. This approach enables a systematic characterization of the main STOV properties in an explicit and physically transparent form. In particular, we analyze the STOV amplitude and phase distributions, their evolution upon free propagation and in optical systems, internal energy flows and the orbital angular momentum. The topologically determined inherent asymmetry of the STOVs and the difference between the "energy center" and "probability center" [Phys. Rev. A **107**, L031501 (2023)] are discussed and qualitatively interpreted. Methods for the STOV generation and diagnostics are outlined, and the main properties of non-Gaussian (Bessel-type) STOVs are briefly described. Finally, limitations of the scalar Gaussian model, accepted throughout the whole text, are considered, and possible generalizations are exposed. The whole presentation may be useful as initial introduction to the STOV-associated ideas and their extraordinary properties.

## I. INTRODUCTION

Short and ultrashort light pulses are important objects of modern optics attracting growing attention of the community. It is these sorts of optical fields with which the main advances and vivid research developments of recent years are associated.[1,2] The especially strong interest is excited to the fields with essential (3+1)-dimensional nature, which are not only characterized by the extremely rapid evolution of their spatial exterior but different parts of the field evolve differently, at the same time exhibiting the deep connections between their evolution in spatial and temporal dimensions – the spatio-temporal (ST) light fields.[2–4] Among the numerous classes and varieties of ST light, the spatio-temporal optical vortices (STOVs) attract a peculiar interest due to the meaningful and non-trivial combination of the expressive ST behavior with the topologically-stipulated optical singularities.[5–8] As a consequence, the STOVs are endowed with exceptional and impressive physical properties, which lead to expectation of striking potential applications.[9,10,11,12,13,14,15,16,17]

Quite naturally, the topic of STOV and associated issues attract the attention of any curious physicist dealing with physical optics, and this text is aimed to facilitate the initial acquaintance with the current problems and prospects of this vigorous field of research. Actually, this text represents a working summary of the current STOV-associated literature, with the special goal of suitable classification and systematization of the related knowledges. It can be considered as a sort of tutorial where the main properties of STOVs as well as principles of their mathematical description and analysis are presented in the form that seems to be most appropriate for persons with good theoretical background in the light-beam optics and look for the consistent, but free of secondary technical details, introduction to the STOV physics. In this context, the main physical principles are discussed based on the simplest examples, and the inevitable complications, associated with their practical realizations, are only briefly mentioned; many practically essential but secondary, in principle, details are omitted.

In this paper, the STOV properties are mainly discussed on the mathematical ground based on the model of paraxial linearly-polarized (scalar) Gaussian wave packet propagating in free space (without dispersion). This way enables a simple interpretation of the STOVs of arbitrary symmetry and arbitrary orders as superpositions of the well-known Hermite-Gaussian modes[17] (Section III, Appendix A). Like in cases of other ST fields, the spatial and temporal Fourier-spectral representations play important role in the STOV theory and practice, and the corresponding details are briefly outlined in Section IV. In particular, these enable to demonstrate the spatially-inhomogeneous spectral composition of STOV fields, the meaningful difference between their "energy center" and "probability center" as manifestations of the inherent lack of their spatial symmetry (some additional details are considered in Appendices B and C).

The angular and temporal spectral densities are involved as derivatives which, however, can be easily realized in practice to perform the necessary transformations and/or for preparing the STOV with desired structure. However, description in the coordinate and time spaces is preferentially used in the whole text. It supplies the direct and physically consistent characterization of the main STOV properties: the amplitude and phase distributions, their evolution during the STOV propagation, associated internal energy flows and the transverse orbital angular momentum[18] (Section V). Situations of combined (transverse + longitudinal) optical vortices (arbitrarily oriented STOV) are considered in Section VI. The analytical description of the propagating Gaussian STOVs of arbitrary orders is presented in Section VII. Problems of the STOV generation, as well as alternative STOV structures, for example, Bessel STOVs, are mentioned briefly in the corresponding Sections VIII and IX. The final Section X supplies a short summary, outlines the simplifications and restrictions accepted for this presentation as well as ways for their overcoming, and possible directions of further research development and application prospects.

**II. SPATIO-TEMPORAL LIGHT FIELDS AND THEIR DESCRIPTION**

We consider the light pulse propagating in free space along the axis $z$; the transverse plane is furnished with the coordinate frame $(x, y)$ As the pulse propagates "as a whole", its field depends on the variable

$$s = z - ct. \tag{1}$$

Instantaneous electric field is described by the wave equation

$$\nabla^2 \boldsymbol{E} - \frac{1}{c^2}\frac{\partial^2 \boldsymbol{E}}{\partial t^2} = 0.$$

In the paraxial approximation,[19,20,21] the electric field is sought in the form

$$\boldsymbol{E} = \mathrm{Re}\left[\mathbf{E}(x,y,z,s)\exp(-i\omega_0 t)\right], \quad \mathbf{E}(x,y,z,s) = \mathbf{u}(x,y,z,s)\exp(ik_0 z) \tag{2}$$

where $k_0 = \omega_0/c$ is the central wavenumber, $\omega_0$ is the central frequency of the wave-packet, and $\mathbf{u}$ is the complex amplitude slowly varying at distances of the order of wavelength $\lambda = 2\pi/k_0$. The complex amplitude can be represented as a series[20]

$$\mathbf{u}(x,y,z,s) = \mathbf{u}^{(0)}(x,y,z,s) + \varepsilon \mathbf{u}^{(1)}(x,y,z,s) + \dots \,. \tag{3}$$

in degrees of the small parameter

$$\varepsilon = (k_0 b)^{-1} \ll 1 \tag{4}$$

where $b$ is the characteristic scale of the field variations in the transverse plane (e.g., the beam width), and $\mathbf{u}^{(0)}$, $\mathbf{u}^{(1)}$, etc., are quantities of the same (zero) order of magnitude.

For simplicity, we consider the linearly polarized field where the zero-order term of expansion (3) is $x$-directed,

$$\mathbf{u}^{(0)}(x, y, z, s) = \mathbf{e}_x u(x, y, z, s). \tag{5}$$

Then, the paraxial equation for $u(x, y, z, s)$ acquires the form[19]

$$\frac{1}{2k_0}\nabla_\perp^2 u + i\frac{\partial u}{\partial z} + \frac{1}{k_0}\frac{\partial^2 u}{\partial z \partial s} = 0 \tag{6}$$

($\nabla_\perp = \mathbf{e}_x(\partial/\partial x) + \mathbf{e}_y(\partial/\partial y)$, $\mathbf{e}_x$, $\mathbf{e}_y$ and $\mathbf{e}_z$ are the unit vectors of the coordinate axes). If the pulse longitudinal inhomogeneity (i.e., its length) is $\zeta$, the last term of Eq. (6) can be evaluated using the relation $\partial/\partial s \simeq 1/\zeta$, which shows that the last term absolute value is $\sim(1/k_0\zeta)$ multiplied by the second term. In usual situations when the longitudinal inhomogeneity contains many wavelengths (the pulse duration $\tau$ contains many, normally $10^3 - 10^4$ periods of the light oscillations), the condition

$$(k_0\zeta)^{-1} \approx (\omega_0\tau)^{-1} \sim \varepsilon \ll 1 \tag{7}$$

takes place (cf. the relation (4)). Consequently, the last term of (6) is of the first order in $\varepsilon$ with respect to other terms, and can be neglected, so that the ST paraxial field can be described by the usual paraxial wave equation[20,21]

$$i\frac{\partial u}{\partial z} = -\frac{1}{2k_0}\nabla_\perp^2 u. \tag{8}$$

Note that this equation does not contain the variable $s$ (1). Accordingly, if a certain function $u_a(x, y, z)$ is a solution to Eq. (8), the function

$$u(x, y, z, s) = F(s) u_a(x, y, z) \tag{9}$$

also satisfies Eq. (8) and describes a physically realizable light field (provided that $F(s)$ is compatible with the common conditions relevant for paraxial fields). For example, we can take the "background" distribution $u_a(x, y, z)$ in the form of Hermite-Gaussian (HG) or Laguerre-Gaussian mode;[21,22] then, a linear combination of different such modes with proper $s$-dependent coefficients $F(s)$ will describe a certain ST wave packet. Note that according to Eq. (9), the argument $z$ of the complex amplitude expresses the "slow" evolution, e.g., dependence of the HG or Laguerre-Gaussian mode parameters on the beam propagation: the "fast" $z$-dependent variations are "captured" by the term $\exp(ik_0 z)$ in (2). (Remarkably, the applicability of Eq. (9) is much broader than the dispersionless free space considered here, and it may provide a good approximation even in the nonlinear case[23]).

According to the paraxial-optics concepts,[20,21] knowledge of the zero-order complex amplitude (5) enables to determine the important field components in the zero and first paraxial orders, and not only the electric $\mathbf{E}$ (2) but also the magnetic $\mathbf{H}$ field vectors:

$$E_x = H_y = u e^{ik_0 s} \sim \varepsilon^0 u, \quad E_z = \frac{i}{k_0}\frac{\partial u}{\partial x} e^{ik_0 s} \sim \varepsilon^1 u, \quad H_z = \frac{i}{k_0}\frac{\partial u}{\partial y} e^{ik_0 s} \sim \varepsilon^1 u \tag{10}$$

(the Gaussian system of electromagnetic units is used). In turn, the momentum density (Poynting vector), averaged over the fast oscillations, can be found from the field components (10) as

$$\mathbf{p} = \frac{1}{8\pi c}\operatorname{Re}\left[\mathbf{E}^* \times \mathbf{H}\right] = \frac{1}{8\pi c}\operatorname{Re}\left[-\mathbf{e}_x E_z^* H_y - \mathbf{e}_y E_x^* H_z + \mathbf{e}_z E_x^* H_y\right]. \tag{11}$$

Explicitly, the momentum-density components are:

$$p_x = -\frac{1}{8\pi c}\mathrm{Re}\left(E_z^* H_y\right) = \frac{1}{8\pi \omega_0}\mathrm{Im}\left(u^* \frac{\partial u}{\partial x}\right), \quad p_y = -\frac{1}{8\pi c}\mathrm{Re}\left(E_x^* H_z\right) = \frac{1}{8\pi \omega_0}\mathrm{Im}\left(u^* \frac{\partial u}{\partial y}\right)$$

$$p_z = \frac{1}{8\pi c}\mathrm{Re}\left(E_x^* H_y\right) = \frac{1}{16\pi c}\left(|E_x|^2 + |H_y|^2\right) = \frac{1}{c}w = \frac{1}{8\pi c}|u|^2 \qquad (12)$$

where $w$ is the energy density of the ST field under consideration. The momentum density **p** determines the energy flow density $c^2\mathbf{p}$. The main energy transfer is associated with the longitudinal component (last expression (12)) which thus characterizes the field intensity distribution; the transverse components $p_x$ and $p_y$ are of the first order in $\varepsilon$ with respect to $p_z$ (see Eqs. (4), (7), (10)) and characterize the internal energy flows.[21]

## III. SPATIO-TEMPORAL OPTICAL VORTICES AND THEIR GENERAL STRUCTURAL FEATURES

The simplest ST optical pulse field of the form (9) is the pulsed Gaussian beam[19] where the Gaussian spatial structure can be expressed by the zero-order HG mode[22,24] $u_{00}^{HG}$ (see Appendix A, Eqs. (A1) – (A3)):

$$u_a(x,y,z) = u_{00}^{HG}(x,y,z) = \frac{A}{b\sqrt{\pi}}\exp\left(-\frac{x^2+y^2}{2b^2} + ik_0\frac{x^2+y^2}{2R} - i\chi\right). \qquad (13)$$

Here we suppose the circular symmetry due to which

$$b_x = b_y = b,\ R_x = R_y = R\ \text{and}\ \chi_x = \chi_y = \chi \qquad (14)$$

(see Eqs. (A2), (A3) in Appendix A). These quantities obey the equations (A2), that is, we suppose the waist cross section to coincide with the plane $z = 0$. In the light pulse, the spatial structure (13) is combined with the Gaussian temporal behavior of $F(s)$ (9):

$$u \equiv u_{ST}^{(0)} = \exp\left(-\frac{s^2}{2\zeta^2}\right)u_{00}^{HG} = \frac{A}{b\sqrt{\pi}}\exp\left(-\frac{x^2+y^2}{2b^2} - \frac{s^2}{2\zeta^2} + ik_0\frac{x^2+y^2}{2R} - i\chi\right). \qquad (15)$$

The normalization constant $A$ determines the field amplitude and is measured in units of electric potential (for example, statV = g$^{1/2}$·cm$^{1/2}$/s). Here, the characteristic scale $b$ of spatial inhomogeneity and the pulse length scale $\zeta$, introduced in Eq. (7), acquire the distinct numerical meanings of the beam half-radius in the transverse ($x$, $y$) cross section and the pulse half-length in the longitudinal ($z$) direction, together with the pulse duration in time $\tau = \zeta/c$.

Equation (15) describes the simplest light pulse with the Gaussian profile in all 3 dimensions. Similarly, the simplest pulse with the transverse optical vortex (OV)[5,21] can be expressed via the general HG modes (A1) as

$$u_{XY}^{(1)} = \frac{1}{\sqrt{2}}\left(u_{01}^{AS} - i\sigma u_{10}^{AS}\right)\exp\left(-\frac{s^2}{2\zeta^2}\right)$$

$$= \frac{A}{\sqrt{\pi b_x b_y}}\left(\frac{y}{b_y}e^{-2i\chi_y} - i\sigma\frac{x}{b_x}e^{-2i\chi_x}\right)\exp\left[-\frac{x^2}{2b_x^2} - \frac{y^2}{2b_y^2} + \frac{ik_0}{2}\left(\frac{x^2}{R_x} + \frac{y^2}{R_y}\right) - \frac{s^2}{2\zeta^2}\right]. \qquad (16)$$

Here, a generally astigmatic transverse OV is presented[25,26] which reduces to the circularly symmetric OV[5,21] upon conditions (14):

$$u_{XY}^{(1)} = \frac{A}{b^2\sqrt{\pi}}(y - i\sigma x)\exp\left[-\frac{x^2+y^2}{2b^2} + \frac{ik}{2R}(x^2+y^2) - 2i\chi\right] \qquad (17)$$

(cf. Eqs. (A1), (A2) for further references). Equations (16) and (17) describe the well-known transverse OVs which are occasionally mentioned in this work mainly as a base for comparisons and instructive analogies. In further consideration we will refer to these OVs as "conventional" OVs, in contrast to the STOVs, which are the main subjects of further narration.

The simplest STOV field can be constructed[17] similarly to (16), (17) as a superposition of two solutions of the form (9) based on the symmetric HG modes with indices (0, 0) and (1, 0):

$$u_{ST}^{(1)} = F_{00}^{(1)}(s) u_{00}^{HG} + F_{10}^{(1)}(s) u_{10}^{HG} \tag{18}$$

where

$$F_{00}^{(1)} = \frac{s}{\zeta} \exp\left(-\frac{s^2}{2\zeta^2}\right), \quad F_{10}^{(1)} = i\frac{\sigma}{\sqrt{2}} \exp\left(-\frac{s^2}{2\zeta^2}\right), \tag{19}$$

$u_{00}^{HG}$ is given by (13) and $u_{10}^{HG}$ is described by Eqs. (A1), (A2) with $m = 1$, $n = 0$, $b_{0x} = b_{0y} = b_0$:

$$u_{10}^{HG} = A\sqrt{\frac{2}{\pi}} \frac{x}{b^2} \exp\left(-\frac{x^2+y^2}{2b^2} + ik_0\frac{x^2+y^2}{2R} - 2i\chi\right). \tag{20}$$

In Eq. (19), the constant $\sigma = \pm 1$ denotes the STOV sign; the normalization constant $A$ is the same as in Eq. (13).

With account for (20) and (13), the superposition (18) can be presented in the more explicit form

$$u_{ST}^{(1)} = \left(\frac{s}{\zeta} u_{00}^{HG} + i\frac{\sigma}{\sqrt{2}} u_{10}^{HG}\right) \exp\left(-\frac{s^2}{2\zeta^2}\right) = \left(\frac{s}{\zeta} + i\sigma\frac{x}{b}e^{-i\chi}\right) u_{ST}^{(0)}$$

$$= \frac{A}{b\sqrt{\pi}}\left(\frac{s}{\zeta} + i\sigma\frac{x}{b}e^{-i\chi}\right) \exp\left(-\frac{x^2+y^2}{2b^2} - \frac{s^2}{2\zeta^2} + ik_0\frac{x^2+y^2}{2R} - i\chi\right) \tag{21}$$

where $u_{ST}^{(0)}$ is given by (15).

Now consider some features of the field (21) in more detail. The intensity distribution (last Eq. (12)) is proportional to $\left|u_{ST}^{(1)}\right|^2$; its section by the $(x, s)$ plane, calculated for the case $\zeta = b_0$, $\sigma = +1$ and $z = 0$, is presented in Fig. 1(a). Remarkably, the distribution is ring-like, with the circular bright ring and the dark spot in the center. Moreover, the isolated amplitude zero in point ($s = 0$, $x = 0$) is coupled with the phase singularity: the field phase is indeterminate at $s = 0$, $x = 0$, and grows by $2\pi$ upon the circulation near this point (Fig. 1(b)). Here, the phase grows on the counter-clockwise circulation; this is associated with the positive sign $\sigma = +1$ of the STOV topological charge. These features resemble the field pattern of a conventional circular OV beam in the transverse $(x, y)$ plane[5] depicted in Fig. 1(c) for comparison.

The 3D spatial profile (intensity distribution) of an OV field forms, generally, a toroidal structure that can be illustrated by the equal-intensity surfaces (Fig. 1(d), (e)). For the STOV (21), the toroid is situated along the longitudinal plane $(s, x)$ containing the propagation axis,[13] while a light pulse with the conventional transverse OV forms a similar toroid in the $(x, y)$ plane (illustrated by Fig. 1(d)); the "depth" of the latter toroid (its size along the $s$-direction) is determined by the pulse duration.

Together with the intensity, Fig. 1(a) presents the distribution of the transverse component of the Poynting vector (11) calculated for the field (18) – (19) (cyan arrows). It shows a sort of circulatory energy flow associated with the corresponding orbital angular momentum (OAM) of the STOV,[13,18,27,28] which also resembles the OAM of a conventional OV[5,21] (Fig. 1(d)) but is directed orthogonally to the beam propagation (Fig. 1(e)). In more detail, the OAM is discussed in Section V; here we note that the internal energy flows can be analytically described by the energy and momentum distributions (11), (12) which for the field (21) read

$$p_z = \frac{1}{c}w = \frac{|A|^2}{8\pi^2 b^2 c}\left[\left(\frac{s}{\zeta}\right)^2 + \left(\frac{x}{b}\right)^2 + 2\sigma\frac{sx}{\zeta b}\sin\chi\right]I_{ST}^{(0)}. \qquad (22)$$

$$p_x = \frac{|A|^2}{8\pi^2 b^2 \omega_0}\left[\sigma\frac{s}{b\zeta}\left(\cos\chi + 2\frac{k_0 x^2}{R}\sin\chi\right) + \frac{k_0 x}{R}\left(\frac{s^2}{\zeta^2} + \frac{x^2}{b^2}\right)\right]I_{ST}^{(0)}$$

$$= \frac{|A|^2}{8\pi^2 b^2 \omega_0}\left\{\sigma\frac{s}{\zeta}\frac{b_0}{b^2}\left[1 + \frac{2x^2}{b_0^2}\left(1 - \frac{b_0^2}{b^2}\right)\right] + \frac{k_0 x}{R}\left(\frac{s^2}{\zeta^2} + \frac{x^2}{b^2}\right)\right\}I_{ST}^{(0)}, \quad p_y = \frac{y}{R}p_z \qquad (23)$$

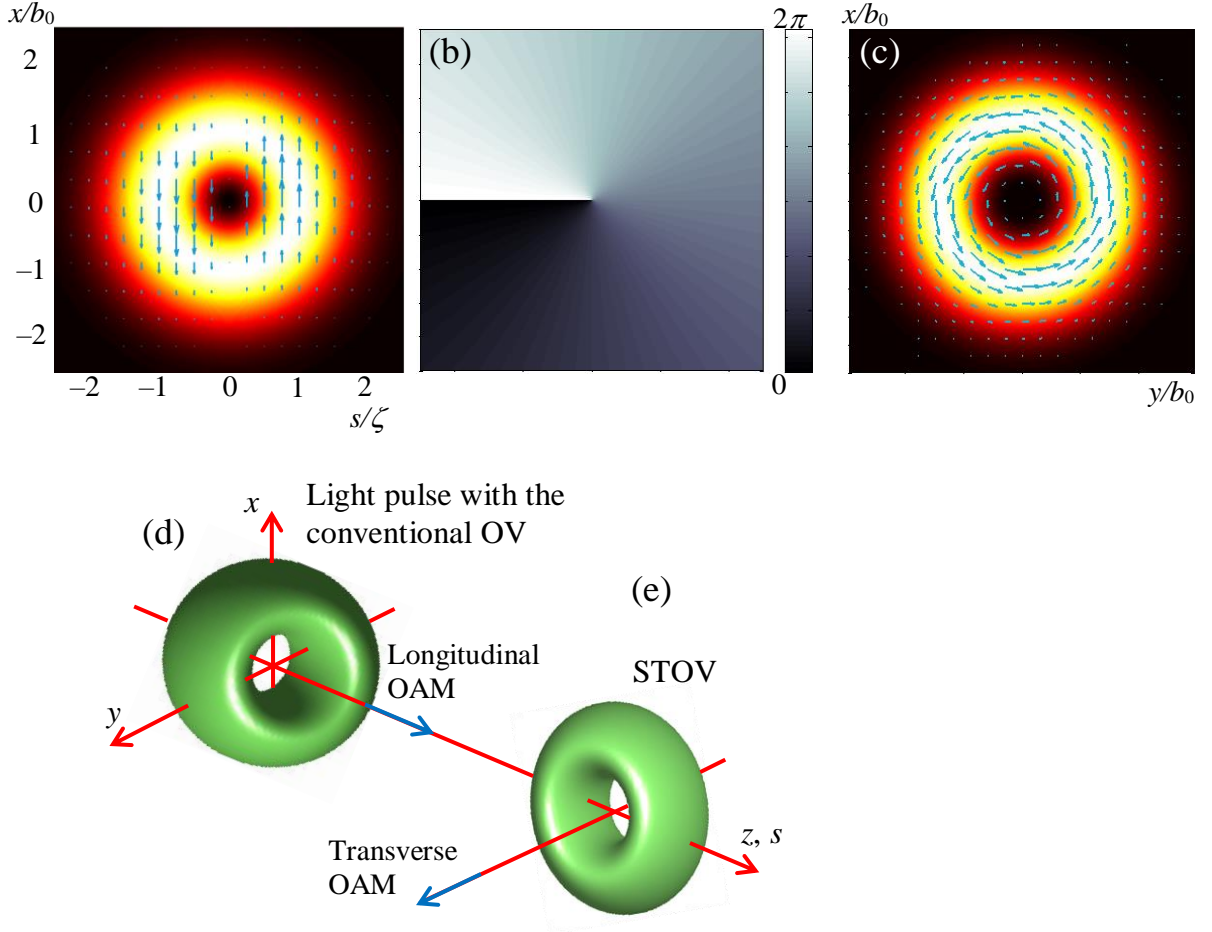

Fig. 1. Characteristics of the STOV (21) with $b_0 = \zeta = 0.1$ mm, $k_0 = 10^5$ cm$^{-1}$, $z = 0$, $\sigma = +1$. (a) Intensity distribution in the plane $(s, x)$, arrows indicate the energy-flow lines (23); (b) phase distribution in the plane $(s, x)$; (c) transverse profile of the symmetric transverse OV described by Eqs. (16), (17), (14). Bottom row represents a comparison between the (d) optical pulse with the conventional transverse OV carrying the longitudinal OAM and (e) the STOV carrying the transverse OAM: green tori are the surfaces at which the intensity is 0.5 of maximum.

where $I_{ST}^{(0)} = \exp\left(-\frac{x^2 + y^2}{b^2} - \frac{s^2}{\zeta^2}\right)$ is the dimensionless intensity "envelope" of the Gaussian ST packet (15). In particular, Eq. (22) enables to calculate the total energy $W$ and the total $z$-directed momentum $P_z$ of the wave packet determined by integration of the energy density $w$ over the infinite ranges of all variables,

$$W = cP_z = \int w(x,y,z,s) dxdyds = \frac{1}{8\pi} \int \left|u_{ST}^{(1)}\right|^2 dxdyds = \frac{|A|^2}{8\sqrt{\pi}} \zeta . \qquad (24)$$

In contrast to the conventional circular OV[5,21] (Fig. 1(c)), in the STOV, a certain "disbalance" exists in the transverse energy flows between the regions $s > 0$ and $s < 0$, which is well seen in Fig. 1(a) and Eq. (23). Due to this disbalance, the spatial configuration of the STOV does not preserve the circular symmetry and changes in the course of propagation. The examples are presented in Fig. 2.

There, the evolution of the light pulse (21) propagating in positive $z$-direction is shown. The "perfect" circular (more exactly, elliptical, as $\zeta$ may generally differ from $b(z)$ at any $z$) STOV structure can only be seen in a single cross section. For the parameters accepted in Figs. 1 and 2, this section is the waist section of the HG beams (A1), (A2). Before the waist ($z < 0$), the OV is anisotropic ("non-canonical");[25,26,30,31] the intensity distribution is asymmetric with two lobes shifted up and down with respect to the propagation axis $z$ (see the images of Fig. 2(a1), (b1) and Fig. 2(a2), (b2) corresponding to $z = -z_R$ and $z = -0.5z_R$). However, the transverse energy flows (arrows) try to redistribute the energy, moving the lobes towards the axis and, ultimately, produce the symmetric distribution at $z = 0$ (Fig. 2(c1), (c2)). Upon further propagation, the energy-flow disbalance continues to distort the intensity profile, and, again, the two lobes appear. The whole pattern behind the waist looks symmetrically to that observed before the waist (images Fig. 2(d1), (e1) and Fig. 2(d2), (e2)). The whole evolution of the propagating STOV looks as a sort of rotation. The sense of this rotation is dictated by the sign of the STOV topological charge which is positive in the upper row of Fig. 2. In case of negative topological charge ($\sigma = -1$), the asymmetric part of the transverse energy flow $c^2 p_x$ (23) inverts, and the intensity pattern of the propagating light pulse rotates oppositely (bottom row of Fig. 2).

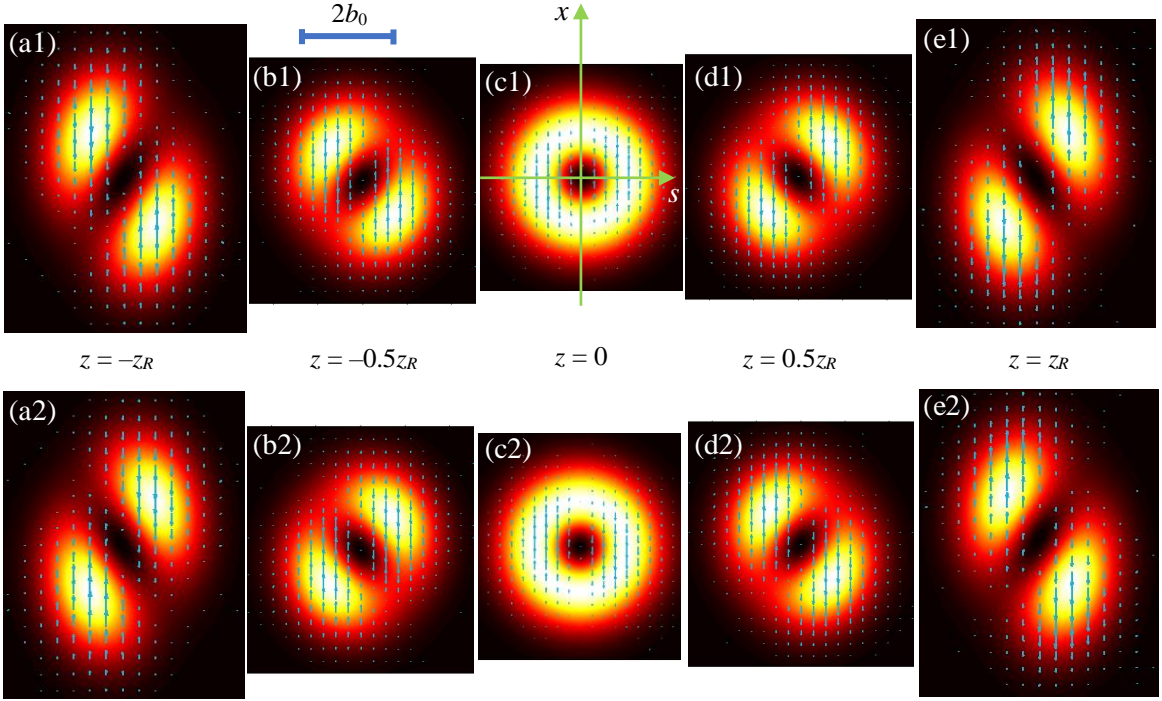

Fig. 2. Spatial evolution of the STOV (21) during propagation: (upper row) $\sigma = 1$, (bottom row) $\sigma = -1$. Other parameters are the same as in Fig. 1; the common scale of the images is indicated above (b1).

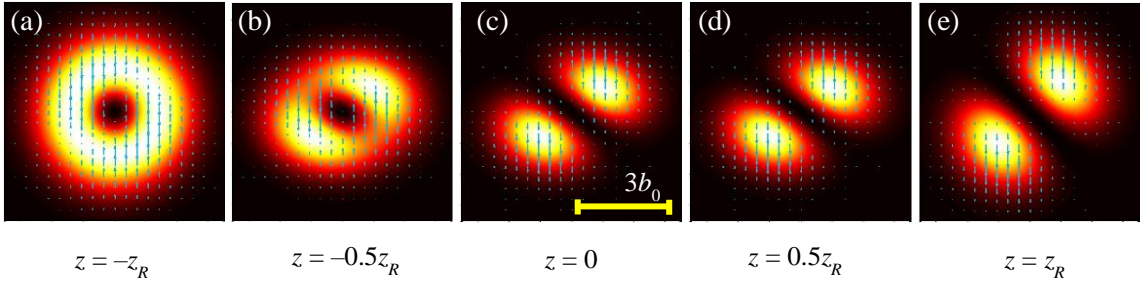

| (a) $z = -z_R$ | (b) $z = -0.5z_R$ | (c) $z = 0$ | (d) $z = 0.5z_R$ | (e) $z = z_R$ |

Fig. 3. Propagation of the STOV (25) – (27) with the "perfect" circular structure at $z_c = -z_R$. The common scale of the images is indicated in (c).

In the examples of Figs. 1, 2, the "perfect" elliptic STOV structure occurs in the waist cross section $z = 0$ of the HG beams (A1), (A2). This is a consequence of the choice $\zeta = b_0$ in Eq. (21). If we want to construct a STOV with the perfect structure in an arbitrary cross section $z = z_c$, we choose the complex value of the constant $\zeta$:

$$\zeta_c = b_c e^{i\chi_c}, \quad \chi_c = \chi(z_c) \tag{25}$$

with real $b_c$, and take the solution of Eq. (8) in the form

$$\left[u_{ST}^{(1)}\right]_f = \left(\frac{s}{\zeta_c} + i\sigma \frac{x}{b} e^{-i\chi}\right) \left[u_{ST}^{(0)}\right]_f \tag{26}$$

where

$$\left[u_{ST}^{(0)}\right]_f = \frac{A}{b\sqrt{\pi}} \exp\left(-\frac{x^2 + y^2}{2b^2} - \frac{s^2}{2|\zeta_c|^2} + ik_0 \frac{x^2 + y^2}{2R} - i\chi\right) \tag{27}$$

(cf. Eq. (21)). At $z = z_c$, this function describes a perfect elliptical profile which becomes circular if $b_c = b(z_c)$ dictated by equations (A2), (14). The corresponding STOV evolution is shown in Fig. 3.

For the conditions (25) – (27), the longitudinal symmetry with respect to the waist cross section is broken; the perfect STOV is realized in the non-waist section $z = z_c$ (in Fig. 3, $z_c = -z_R$ is chosen). The transformations presented in Fig. 3 can be interpreted in terms of the STOV focusing. Initially, a "perfect" STOV exists with the intensity distribution of Fig. 3(a) and the planar wavefront (waist) at $z = z_c$ but at this cross section, the wavefront curvature is introduced (3$^{rd}$ summand in parentheses of Eq. (27) can be treated as a result of passage through a thin lens with focal distance $f = -R(z_c)$). Then, a focused STOV is formed whose waist cross section $z = 0$ occurs in the plane of focusing. Fig. 3 testifies that in this plane (as well as in any other plane, except the initial one $z = z_c$), the STOV shape is not "perfect" but tends to a two-lobe structure.

## IV. SPATIAL AND TEMPORAL FOURIER-SPECTRAL REPRESENTATIONS

Fourier transformations of optical fields are widely used in optics because they can be easily realized by simple optical means: angular spectrum is visualized by the far-field distribution or in the focal plane of a lens, the frequency (wavelength) spectrum is mapped into a spatial pattern by gratings or other dispersive elements. That is why, in all manipulations with structured light fields, the Fourier-conjugated domains are equally available and can be equally employed; the only question is which domain offers more simple and feasible means for the field handling. For the ST fields, the possibilities associated with the Fourier transformations are especially profitable (see also Section VIII).

Generally, the "full" Fourier-spectrum of a ST field includes transformations over all spatial and temporal variables, and in the paraxial case (2), (6), it is described by expression

$$U(k_x, k_y, k) = \int u(x, y, s) \exp\left[-i(k_x x + k_y y) - is\Delta k\right] dx dy ds \tag{28}$$

where $\Delta k = k - k_0$, $k = \omega/c$ is the wavenumber corresponding to the spectral frequency $\omega$, and $u(x, y, s)$ is determined by (9). Here, we conventionally consider the STOV characteristics at a certain fixed longitudinal position $z =$ const, and the fourth argument $z$ is omitted. The energy density in the $k$-space is described by the function

$$w(k_x, k_y, k) = \frac{1}{8\pi} |U(k_x, k_y, k)|^2 \tag{29}$$

so that the total energy of the wave packet equals

$$W = \frac{1}{8\pi} \frac{1}{(2\pi)^3} \int |U(k_x, k_y, k)|^2 dk_x dk_y dk . \tag{30}$$

For example, the Fourier spectrum of the STOV field (21) reads

$$U(k_x, k_y, k) = 2^{3/2} \pi A \zeta b_0 (-i\zeta \Delta k + \sigma k_x b_0) \exp\left[-\frac{1}{2}(k_x^2 + k_y^2) b_0^2 - \frac{1}{2} \zeta^2 \Delta k^2\right] \tag{31}$$

(here and further in this Section we address the STOV characteristics in the waist plane $z = 0$, $b = b_0$). One may persuade that, in view of the Parseval formula,[32] the result (30) coincides with the previously calculated STOV energy (24).

An interesting property of the ST fields is that their spectral composition is, generally, spatially inhomogeneous. This can be suspected from expression (31) but in more direct way is illustrated by the combined distribution in the space and spectral domains, which follows from the partial Fourier transform of (21) with respect to the "temporal" coordinate $s$:

$$U_{ST}^{(1)}(x, y, k) = \int u_{ST}^{(1)}(x, y, s) e^{-is\Delta k} ds = A\sqrt{2} \frac{i\zeta}{b_0}\left(\sigma \frac{x}{b_0} - \zeta \Delta k\right) \exp\left(-\frac{x^2 + y^2}{2b_0^2} - \frac{\zeta^2}{2} \Delta k^2\right). \tag{32}$$

The function $|U_{ST}^{(1)}(x, y, k)|^2$ describes the energy distribution over the spatial and spectral coordinates so that the density of the spectral component with the wavenumber $k$ in a point with coordinates $(x, y)$ is proportional to $|U_{ST}^{(1)}(x, y, k)|^2$. Fig. 4 demonstrates the section of this function by the plane $y = 0$, calculated under the same conditions as were accepted in Fig. 1. It is seen that at $\sigma = +1$ (Fig. 4(a)), spectral components with $\Delta k < 0$ prevail in the upper half-space ($x > 0$) but are deficient in the lower half-space ($x < 0$). Numerically this effect can be expressed by the $x$-dependent mean value of the spectral shift:

$$\overline{\Delta k}(x) = \frac{\int \Delta k |U_{ST}^{(1)}(x, y, k)|^2 dk}{\int |U_{ST}^{(1)}(x, y, k)|^2 dk} = -\frac{\sigma}{b_0 \zeta} \frac{2x}{1 + 2x^2/b_0^2}. \tag{33}$$

Note that the asymmetry of the images in Fig. 4 is closely associated with the vortex nature of the STOV (helical phase distribution in the $(s, x)$ plane): if the multiplier $i\sigma$ in the first parentheses of $u_{ST}^{(1)}(x, y, s)$ (21) is replaced by any real number, the corresponding spectral density $|U_{ST}^{(1)}(x, y, k)|^2$ would be an even function of $x$.

Physically, the spectral disbalance of the upper ($x > 0$) and lower ($x < 0$) half spaces stems from the vortex phase of the STOV (21). For a propagating STOV, the phase increment between the planes, say, $s = -s_0$ and $s = +s_0$ ($s_0 \gg \zeta$), contains not only the "background" phase difference $2ik_0 s_0$ (see Eq. (2)) but also the vortex-phase contribution proportional to $-\sigma x/|x|$. This corresponds to effective increase of the wavelength in the region where $\sigma x/|x| > 0$ (see Appendix C for details).

The spatial asymmetry of the spectral composition entails another important feature of the STOV fields. For example, in the situation of Fig. 4(a) ($\sigma = +1$), predominance of lower-frequency components in the region $x > 0$ means that photons of lower energies are there accumulated, while the region $x < 0$ is, oppositely, enriched by the higher-energy photons. At the same time, the energy (intensity) profile of the whole field in the $(s, x)$ coordinates is symmetric (see Fig. 1(a)): the upper and lower half-spaces contain equal amounts of energy. Consequently, the half-space $x > 0$ where the photons have, on the average, lower energies, should contain more photons than the region $x < 0$ where photons of higher energies are gathered. Of course, if the STOV topological charge reverses ($\sigma = -1$, Fig. 4(b)), the picture is quite opposite.

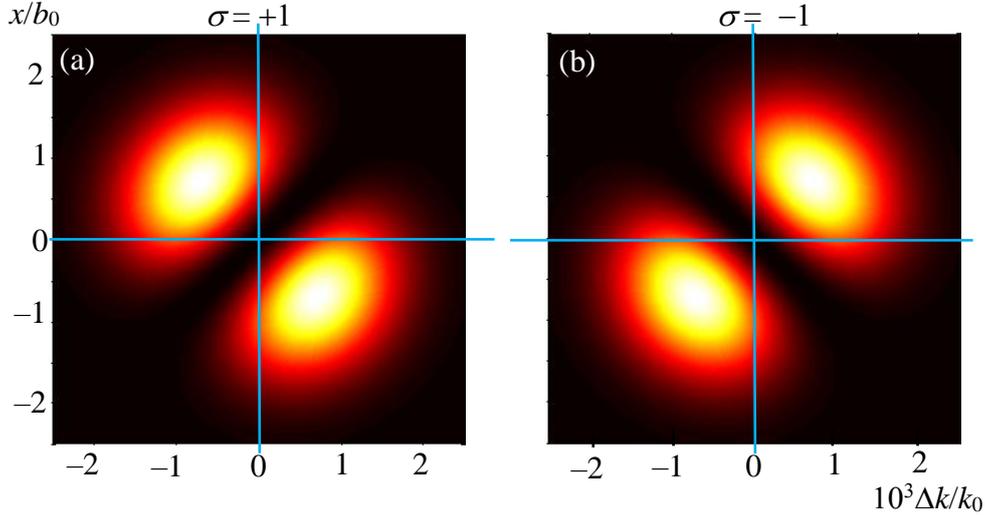

Fig. 4. Color-code view of the distribution $\left|U_{ST}^{(1)}(x,0,k)\right|^2$ (see Eq. (32)) for the STOV of Eqs. (21) in the waist cross section $z = 1/R = \chi = 0$, $b = b_0$, with parameters $\zeta = b_0 = 0.1$ mm, $k_0 = 10^5$ cm$^{-1}$; (a) $\sigma = +1$, (b) $\sigma = -1$.

The effects of photons' redistribution can be described by the photon density $w_p$ (also called "probability distribution"), which can be derived from the energy density $w$ (see, e.g., Eq. (22)) in a regular way as[28,29,34]

$$w_p = \frac{w}{\hbar \omega} = \frac{1}{c\hbar}\frac{w}{k}. \qquad (34)$$

However, the photon-density distribution is ill-defined in the coordinate space,[29] and application of Eq. (34) to the spatial energy density (24) requires special precautions. But it is fully applicable to the energy density in $k$-space where $w \to w(k_x, k_y, k)$ determined by Eq. (29); for example, the total number of photons in the considered wave packet is determined by

$$N_p = \frac{1}{\hbar c}W_p, \quad W_p = \int \frac{1}{k}\left|U(k_x,k_y,k)\right|^2 dk_x dk_y dk. \qquad (35)$$

In particular, the inhomogeneous probability density can be characterized by the "probability center" (PC) position.[28,29] It is determined akin to the energy center that can be expressed via (22) and (24) as

$$\begin{pmatrix} s_e \\ x_e \end{pmatrix} = \frac{1}{W} \int \begin{pmatrix} s \\ x \end{pmatrix} w(x,y,s) dx dy ds. \qquad (36)$$

For the STOV (21) this equation gives

$$s_e = x_e = 0 \qquad (37)$$

which expectedly coincides with the coordinate origin. The similar expression for the PC involves the $k$-domain amplitude (28) and energy density (29), and defines the PC as the weighted mean value of the position operator $i\nabla_k$ in $k$-space:[28,29]

$$\begin{pmatrix} s_p \\ x_p \end{pmatrix} = \frac{1}{W_p} \int \frac{1}{\sqrt{k}} U^*(k_x,k_y,k) i \begin{pmatrix} \partial/\partial k \\ \partial/\partial k_x \end{pmatrix} \left[ \frac{1}{\sqrt{k}} U(k_x,k_y,k) \right] dk_x dk_y dk \qquad (38)$$

(in contrast to the prototype,[29] this expression explicitly highlights the Hilbert-space vectors in the form $k^{-1/2} U(k_x,k_y,k)$ and warrants the real values of $(s_p, x_p)$).

In particular, by using the identity

$$U(k_x,k_y,k) = \int U_{\text{ST}}^{(1)}(x,y,k) e^{-i(k_x x + k_y y)} dx dy$$

where $U_{\text{ST}}^{(1)}(x,y,k)$ is the partial spectral density (32), Eq. (38) gives

$$x_p = \frac{1}{W_p^{xy}} \int \frac{x}{k} \left| U_{\text{ST}}^{(1)}(x,y,k) \right|^2 dx dy dk, \quad W_p^{xy} = \int \frac{1}{k} \left| U_{\text{ST}}^{(1)}(x,y,k) \right|^2 dx dy dk. \qquad (39)$$

Note that the same expression for the PC coordinate can be obtained directly from the analog of Eq. (36) where the ST energy density $w(x,y,s)$ is replaced by $\left| U_{\text{ST}}^{(1)}(x,y,k) \right|^2$ – the energy distribution over the spatial $(x,y)$ and spectral $(k)$ coordinates, with the corresponding modification of the integration domain. The value of $x_p$ (39) can be found explicitly by using the condition (7) due to which the spectral width of the STOV pulse is much less than the central frequency. In this situation, the inequality $\Delta k << k_0$ holds for the physically relevant range of $k$, which enables the approximation

$$\frac{1}{k} \simeq \frac{1}{k_0} \left( 1 - \frac{\Delta k}{k_0} \right) \qquad (40)$$

in the above integrals. Accordingly, with employment of the explicit expression (32), Eqs. (39) yield

$$x_p = \frac{\sigma}{2k_0} \frac{b_0}{\zeta} = \frac{\sigma}{2k_0 \gamma} \qquad (41)$$

where $\gamma = \zeta/b_0$ is the parameter of the STOV ellipticity (anisotropy).[25,26,28,29] Similarly, it can be shown that $s_p = 0$ (Appendix B).

In agreement with the note below Eq. (33), the result (41) reflects the deep topological nature of the STOV. It supplies a quantitative expression of the photon-number disbalance between the $x > 0$ and $x < 0$ half-spaces, mentioned above in the comment to Fig. 4. As a general conclusion, Eq. (41) indicates that the STOV structure cannot be perfectly symmetric.[29] It contains at least two physically meaningful centers that can be considered as its generalized spatial markers: the energy center and the PC. Their roles will be further discussed, e.g., for the definition and properties of the STOV-associated OAM.

## V. ORBITAL ANGULAR MOMENTUM OF SPATIO-TEMPORAL OPTICAL VORTICES

It was noted in the previous Section III that the specific distribution of the Poynting vector in the plane ($x$, $s$) is coupled with the *transverse* OAM with respect to any $y$-oriented axis. It is suitable to consider the OAM defined with respect to the moving axis ($x = 0$, $s = 0$) crossing the pulse center. Thus, the $y$-component of the Poynting vector (11) gives no contribution, and the OAM density can be determined as

$$L_y = sp_x - xp_z. \tag{42}$$

The total OAM of the STOV is obtained via the integration of (42) over $dxdyds$. In this procedure, the term $xP_z$ gives a zero contribution due to the symmetry of expression (22), and, in view of (23), one obtains for the total OAM

$$\Lambda_y = \int sp_x\, dxdyds$$
$$= \frac{|A|^2 b_0}{8\pi^2 \omega_0 b^4} \frac{\sigma}{\zeta} \int s^2 \left[1 + \frac{2x^2}{b_0^2}\left(1 - \frac{b_0^2}{b^2}\right)\right] \exp\left(-\frac{x^2 + y^2}{b^2} - \frac{s^2}{\zeta^2}\right) dxdyds = \frac{|A|^2 \sigma}{16\sqrt{\pi}\omega_0} \frac{\zeta^2}{b_0}. \tag{43}$$

In agreement with the angular momentum conservation, this result does not depend on $z$, despite that the pulse configuration changes rather impressively (as is seen in Figs. 2, 3). The quantity (43) depends on a number of the STOV parameters; however, like the longitudinal OAM of the conventional OV beams, the transverse OAM (43) expresses the deep topological properties of the field which are largely "masked" by the specific parameters of the beam shape. To disclose this topological essence, the numerical OAM value (43) should be properly normalized. The conventional OAM measure "per photon" is somewhat contradictory in application to essentially polychromatic STOV fields[28,29] but the OAM per unit energy can be used instead. According to Eqs. (43) and (24), the OAM per unit energy of the STOV is[27–29]

$$\frac{\Lambda_y}{W} = \frac{\sigma}{2\omega_0} \frac{\zeta}{b_0} = \frac{\sigma}{2} \frac{\gamma}{\omega_0} \tag{44}$$

where $\gamma$ is the measure of the STOV anisotropy introduced in Eq. (41).

This result can be compared with the longitudinal OAM $\Lambda_{OV}$ of the conventional transverse OV. In view of the generally astigmatic character of the STOV, for comparison we choose the light pulse with the first-order astigmatic transverse OV (16), for which the longitudinal OAM $\Lambda_{OV}$, normalized per unit energy, obeys the relation[29,25,26]

$$\frac{\Lambda_{OV}}{W} = \frac{\sigma}{2\omega_0}\left(\gamma_{xy} + \gamma_{xy}^{-1}\right) \tag{45}$$

where $\gamma_{xy} = b_{0y}/b_{0x}$. Remarkably, in the symmetric case where $\gamma = 1$ in Eq. (44) and $\gamma_{xy} = 1$ in Eq. (45), the simple correspondence takes place:

$$\frac{\Lambda_y}{W} = \frac{\sigma}{2\omega_0} = \frac{1}{2}\frac{\Lambda_{OV}}{W}. \tag{46}$$

This relation was a subject of a controversial discussion[11,17,27,29,35] but it finds a simple qualitative support in juxtaposition of the corresponding energy flow patterns presented in Figs. 1(a) and 1(c).[17,27] The energy circulation in the conventional OV (Fig. 1(c)) is "complete" and contains the contributions along both orthogonal transverse components while in the STOV field, only the $\pm x$-oriented contributions are present; accordingly, the circulation loses a half of its "complete" value.

It should be noted, however, that the OAM of Eqs. (42) – (44), (46), determined with respect to the STOV energy center ($x = 0$, $s = 0$), is not the only meaningful form of the STOV-associated OAM. As was noted at the end of Section IV, the STOV possesses two physically meaningful spatial centers, and the PC (41) can be considered as an alternative marker of the STOV position. Accordingly, the OAM with respect to the PC can be introduced as the "intrinsic" OAM of the STOV[28,29] (and this definition offers certain advantages from the relativistic point of view[18,36]). In contrast to the "absolute" OAM (42), its density is determined by

$$L_y^{\text{int}} = s p_x - (x - x_p) p_z = L_y + x_p p_z. \tag{47}$$

This equation, together with Eq. (44), immediately entails that the total intrinsic OAM of the STOV is described by relation

$$\frac{\Lambda_y^{\text{int}}}{W} = \frac{\Lambda_y}{W} + \frac{x_p}{W} \int p_z dx dy ds = \frac{\Lambda_y}{W} + \frac{\sigma}{2\omega_0} \frac{1}{\gamma} = \frac{\sigma}{2\omega_0}\left(\gamma + \frac{1}{\gamma}\right). \tag{48}$$

Both the OAM (44) with respect to the energy center and the OAM (48) with respect to the PC express the same physical properties. Their interrelation follows from the universal relation uniting the AMs of the same physical object, carrying the momentum **P**, with respect to different reference points **R**$_1$ and **R**$_2$:

$$\mathbf{\Lambda}(\mathbf{R}_1) = \mathbf{\Lambda}(\mathbf{R}_2) - (\mathbf{R}_1 - \mathbf{R}_2) \times \mathbf{P}; \quad \Lambda_y(x_p) - \Lambda_y(0) = \Lambda_y^{\text{int}} - \Lambda_y = x_p P_z.$$

The difference $\Lambda_y^{\text{ext}} = \Lambda_y - \Lambda_y^{\text{int}}$ can be called "extrinsic" OAM;[29] this quantity is of a purely geometrical meaning and equals to

$$\frac{\Lambda_y^{\text{ext}}}{W} = -\frac{\sigma}{2\omega_0}\frac{1}{\gamma}. \tag{49}$$

The extrinsic OAM (49) characterizes the angular momentum (AM) with respect to the origin calculated in supposition that the whole field momentum is applied at the PC point. Due to the condition (41), this AM is always directed oppositely to the "basic" energy circulation, for example, when $x_p > 0$ (condition of Fig. 1), $\Lambda_y^{\text{ext}}$ acts clockwise (is negative). This property is compatible with the fact that the extra number of photons are accumulated at that side of plane $x = 0$, towards which the energy-circulation flow is directed in the frontal part of the STOV.

To conclude this Section, we note that another approach to the OAM of an ST wave packet is possible, in which the OAM is determined as the time-integrated AM flux through a fixed transversal plane,[27,37] rather than the volume-integrated AM density accepted here according to Refs. 11,17,28,29. In essence, both approaches are physically equivalent but involve different details of the ST-field behavior (temporal evolution in a fixed transverse plane vs spatial inhomogeneity in a fixed moment of time). This leads to meaningful peculiarities in the interpretations of the OAM and its "intrinsic" or "extrinsic" constituents[37] which, however, cannot be properly discussed within the limited frame of this paper (see some additional remarks in the 1$^{\text{st}}$ paragraph of Section X).

## VI. ARBITRARILY ORIENTED SPATIO-TEMPORAL OPTICAL VORTICES

The STOVs (21), (26) considered above are oriented such that their intensity toroids and the energy circulation are concluded within the plane $(z, x)$ (see Fig. 1). However, solutions of the ST paraxial equation (6) can be built with other toroid orientations.[15] For example, employing the Hermite-Gaussian functions (A1), one can choose

$$u_{\text{ST}}^{(11)} = \left(\alpha \frac{s}{\zeta} u_{00}^{AS} + \frac{\beta}{\sqrt{2}} u_{01}^{AS} + i\frac{\sigma}{\sqrt{2}} u_{10}^{AS}\right) \exp\left(-\frac{s^2}{2\zeta^2}\right)$$

$$= \frac{A}{\sqrt{\pi b_x b_y}} \left( \alpha \frac{s}{\zeta} + \beta \frac{y}{b_y} e^{-i\chi_y} + i\sigma \frac{x}{b_x} e^{-i\chi_x} \right)$$

$$\times \exp\left[ -\frac{x^2}{2b_x^2} - \frac{y^2}{2b_y^2} - \frac{s^2}{2\zeta^2} + \frac{ik_0}{2}\left(\frac{x^2}{R_x} + \frac{y^2}{R_y}\right) - i\frac{1}{2}(\chi_x + \chi_y) \right]. \quad (50)$$

This wave packet is, again, of the STOV structure, and propagates along axis $z$ but its equal-intensity toroid is adjusted along the plane $(x, u)$ which is inclined at an angle $\varphi$ (Fig. 5). For comparison with the "pure" transverse STOV (21), it is suitable to keep the condition $\alpha^2 + \beta^2 = 1$ in (50); then, $\alpha$ and $\beta$ appear as the direction cosines of the STOV orientation: $\alpha = \cos\varphi$, $\beta = \cos(\pi/2 - \varphi)$. For example, if $b_y = \zeta$, $\chi_y = 0$ (conditions of Fig. 1) and $\alpha = \beta = 1/\sqrt{2}$, $\varphi = \pi/4$. Practically, STOVs of arbitrary orientation can be realized.[15]

The dynamical characteristics of the "oblique" STOV of Eq. (50) are determined similarly to Eqs. (22), (23) but acquire modified forms; for the symmetric case (14) one obtains:

$$p_z = \frac{1}{c} w = \frac{|A|^2}{8\pi^2 b^2 c} \left[ \alpha^2 \left(\frac{s}{\zeta}\right)^2 + \beta^2 \left(\frac{y}{b}\right)^2 + \left(\frac{x}{b}\right)^2 + 2\alpha\beta \frac{sy}{b\zeta}\cos\chi + 2\sigma\alpha \frac{sx}{b\zeta}\sin\chi \right] I_{ST}^{(0)}, \quad (51)$$

$$p_x = \frac{|A|^2}{8\pi^2 b^2 \omega_0} \frac{\sigma}{b} \left( \alpha \frac{s}{\zeta}\cos\chi + \beta \frac{y}{b} \right) I_{ST}^{(0)} + \frac{x}{R} p_z,$$

$$p_y = -\frac{|A|^2}{8\pi^2 b^2 \omega_0} \frac{\beta}{b} \left( \alpha \frac{s}{\zeta}\sin\chi + \sigma \frac{y}{b} \right) I_{ST}^{(0)} + \frac{y}{R} p_z. \quad (52)$$

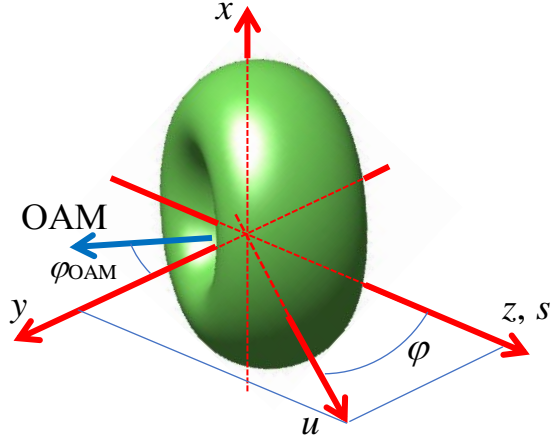

Fig. 5. Illustration of the STOV oriented at an angle $\varphi$ with respect to the propagation axis $z$. The STOV spatial distribution is represented by the equal-intensity toroid corresponding to 0.5 of the maximum.

Accordingly, the total STOV energy is

$$W = cP_z = \frac{1}{8\pi} \int \left|u_{ST}^{(11)}\right|^2 dxdyds = \frac{|A|^2}{8\sqrt{\pi}} \zeta \frac{\alpha^2 + \beta^2 + 1}{2} \xrightarrow{\alpha^2+\beta^2=1} \frac{|A|^2}{8\sqrt{\pi}} \zeta \quad (53)$$

which expectedly coincides with the result (24) for the $(x, z)$-oriented STOV. Combining Eqs. (36) and (51) shows that the energy-center position still obeys Eq. (37).

One can easily find the spectral representations (see Section IV) of the arbitrarily oriented STOV. For example, upon the symmetric conditions (14) and for the fixed longitudinal position $z = 0$, the spectral density (28) appears in the form

$$U^{(11)}(k_x,k_y,k) = 2^{3/2}\pi A\zeta b\left[-i\zeta\alpha\Delta k - i\beta k_y b + \sigma k_x b\right]\exp\left[-\frac{1}{2}\left(k_x^2 + k_y^2\right)b^2 - \frac{1}{2}\zeta^2\Delta k^2\right] \quad (54)$$

(cf. Eq. (31) for the longitudinal STOV). Likewise, the analogue of the partial Fourier transform (32) can be obtained

$$U_{ST}^{(11)}(x,y,k) = \int u_{ST}^{(11)}(x,y,s)e^{-is\Delta k}ds$$
$$= A\sqrt{2}\frac{\zeta}{b_0}\left(i\sigma\frac{x}{b_0} + \beta\frac{y}{b_0} - i\alpha\zeta\Delta k\right)\exp\left(-\frac{x^2+y^2}{2b_0^2} - \frac{\zeta^2}{2}\Delta k^2\right). \quad (55)$$

Just like Eq. (32), this expression reflects the spatial inhomogeneity of the spectral composition; the corresponding analogue of Eq. (33) reads

$$\overline{\Delta k}(x) = -\frac{\sigma}{b_0\zeta}\frac{2\alpha x}{\alpha^2 + 2(x^2 + \beta^2 y^2)/b_0^2}. \quad (56)$$

When $\alpha \to 1$, $\beta \to 0$ this result coincides with (33) while at the opposite conditions $\alpha \to 0$, $\beta \to 1$, $\overline{\Delta k}(x)$ expectedly vanishes because the STOV degenerates into a sort of the conventional transverse OV (16), (17). Via the analogues of Eqs. (35), (39) and using Eq. (40), it is possible to find the PC position for the oblique STOV (50). Under the symmetric conditions (14) and $z = 0$ they give

$$x_p = \sigma\frac{\alpha}{k_0}\frac{b_0}{\zeta}\frac{1}{1+\alpha^2+\beta^2} \xrightarrow{\alpha^2+\beta^2=1} \sigma\frac{\alpha}{2k_0\gamma}, \quad (57)$$

which expectedly vanishes if the STOV orientation approaches the conventional configuration (16), (17) ($\alpha \to 0$).

Operating similarly to (42), (43), we find the OAM of the STOV (50) under conditions (14). In contrast to Section V, now the OAM has two non-zero components:

$$\Lambda_y = \int sp_x\, dxdyds = \sigma\alpha\frac{|A|^2}{8\sqrt{\pi}\,\omega_0}\frac{\zeta^2}{2b_0} = \sigma\alpha\frac{|A|^2}{8\sqrt{\pi}\,\omega_0}\frac{\gamma\zeta}{2}, \quad \frac{\Lambda_y}{W} = \alpha\frac{\sigma}{2\omega_0}\gamma\,;$$

$$\Lambda_z = \int(xp_y - yp_x)dxdyds = -\sigma\beta\frac{|A|^2}{8\sqrt{\pi}\,\omega_0}\zeta, \quad \frac{\Lambda_z}{W} = -\beta\frac{\sigma}{\omega_0}. \quad (58)$$

These equations show that the OAM of the arbitrarily oriented STOV is not obligatory orthogonal to the plane of its orientation: while the STOV plane deviates from the $(x, z)$-plane by an angle $\varphi = \arctan(\beta/\alpha)$, the OAM direction deviates from the $y$-axis by an angle $\varphi_{OAM} = \arctan(2\beta b_0/\alpha\zeta)$ (see Fig. 5). This happens because in the purely longitudinal STOV ($\beta = 0$), the circulation is "incomplete" (only the vertical $\pm x$-oriented contributions are present, see Fig. 1(a)), and with growing $\beta$, the OAM component $\Lambda_z$ emerges not only as a geometric projection of the transverse OAM inherent in the "oblique" STOV (Fig. 5), but also includes the contributions of the horizontal $\pm y$-oriented energy flows (cf. Fig. 1(c)) that are "lost" in the longitudinal STOV (cf. the last paragraph of the previous Section). Accordingly, the absolute value $|\Lambda_z|$ grows faster than is dictated by the STOV inclination, and $\varphi_{OAM} > \varphi$. Only in the marginal situations $\alpha = 0$ and $\beta = 0$ these angles coincide.

The intrinsic OAM of the oblique STOV (50) follows from the proper generalization of Eqs. (47), (48) with account for (57):

$$L_y^{int} = L_y + x_p p_z, \quad L_z^{int} = L_z - x_p p_y;$$

$$\frac{\Lambda_y^{\text{int}}}{W} = \frac{\Lambda_y}{W} + \frac{\sigma}{2\omega_0}\frac{\alpha}{\gamma} = \frac{\sigma\alpha}{2\omega_0}\left(\gamma + \frac{1}{\gamma}\right); \quad \Lambda_z^{\text{int}} = \Lambda_z. \tag{59}$$

These equations confirm that the OAM (58) of the oblique STOV possesses a "mixed" physical nature combining the two parts. The component $\Lambda_y$ is formed as a projection of the "inherent" STOV-associated transverse OAM (43), (44) and shows the same physical regularities, viz. the specific manifestations sensitive to the difference between energy and probability distributions. In contrast, the component $\Lambda_z$ appears as a projection of the "traditional" longitudinal OAM of the conventional OV (16), (17), for which the difference between the energy and probability distributions entails no physical consequences.

**VII. SPATIO-TEMPORAL OPTICAL VORTICES OF HIGHER ORDERS**

So far, we considered the single-charge STOVs for which the phase increment on a round trip near the STOV center ($x = 0$, $s = 0$) is $\pm 2\pi$ (Fig. 1(b)). However, like in case of conventional transverse OVs, the higher-order STOVs can exist for which the phase growth is $2\pi l$, with arbitrary integer topological charge $l$. Their expressions can be "built" from the HG modes (A1), (A2) similarly to Eq. (18) for the case $l = 1$. Accordingly, for the STOV of the order $|l| = 2$, the complex amplitude distribution is derived in the form

$$u_{\text{ST}}^{(2)} \propto F_{00}^{(2)}(s)u_{00}^{HG} + F_{10}^{(2)}(s)u_{10}^{HG} + F_{20}^{(2)}(s)u_{20}^{HG} \tag{60}$$

where

$$F_{00}^{(2)} = \left[\left(\frac{s}{\zeta}\right)^2 - \frac{1}{2}\right]\exp\left(-\frac{s^2}{2\zeta^2}\right), \quad F_{10}^{(2)} = i\sigma\sqrt{2}\frac{s}{\zeta}\exp\left(-\frac{s^2}{2\zeta^2}\right),$$

$$F_{20}^{(2)} = -\frac{1}{\sqrt{2}}\exp\left(-\frac{s^2}{2\zeta^2}\right).$$

Finally,

$$u_{\text{ST}}^{(2)} = \frac{1}{2}\left[\left(\frac{s}{\zeta}\right)^2 - \frac{1}{2} + 2i\sigma\frac{sx}{\zeta b}e^{-i\chi} - \left(\frac{x^2}{b^2} - \frac{1}{2}\right)e^{-2i\chi}\right]u_{\text{ST}}^{(0)} \tag{61}$$

where $u_{\text{ST}}^{(0)}$ is given by Eq. (15), and the multiplier 1/2 is introduced to maintain the STOV energy normalization (24) for $u_{\text{ST}}^{(2)}$. The STOV charge sign is determined by $\sigma = \pm 1$. The evolution of this STOV is illustrated by Fig. 6.

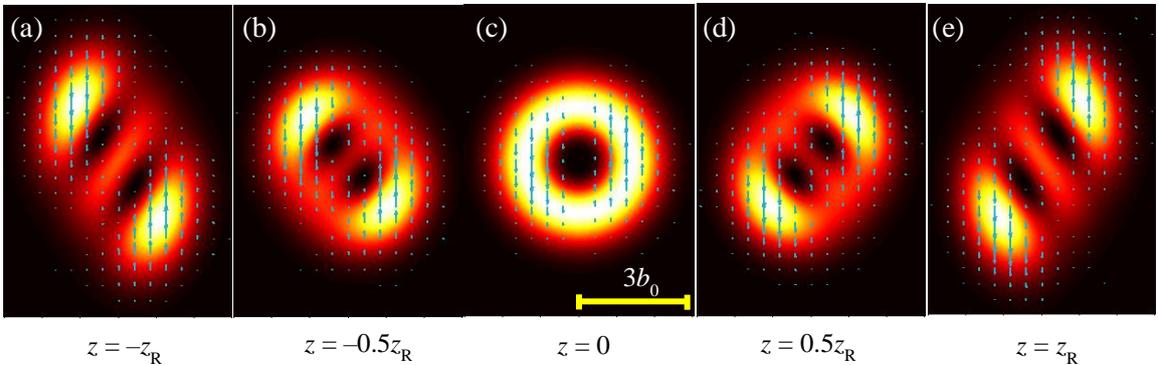

$z = -z_R$     $z = -0.5z_R$     $z = 0$     $z = 0.5z_R$     $z = z_R$

Fig. 6. Evolution of the STOV (60) – (61) with $\sigma = 1$ during propagation; images (a) – (e) differ by the longitudinal position of the propagating STOV (marked below the images). The spatial parameters are the same as in Fig. 1; the common scale of the images is indicated in (c).

Similarly, for the STOV of the order $|l| = 3$,
$$u_{ST}^{(3)} \propto F_{00}^{(3)}(s) u_{00}^{HG} + F_{10}^{(3)}(s) u_{10}^{HG} + F_{20}^{(3)}(s) u_{20}^{HG} + F_{30}^{(3)}(s) u_{30}^{HG} \tag{62}$$

where
$$F_{00}^{(3)} = \left[\left(\frac{s}{\zeta}\right)^3 - \frac{3}{2}\frac{s}{\zeta}\right]\exp\left(-\frac{s^2}{2\zeta^2}\right), \quad F_{10}^{(3)} = \frac{3}{\sqrt{2}}\sigma i\left[\left(\frac{s}{\zeta}\right)^2 - \frac{1}{2}\right]\exp\left(-\frac{s^2}{2\zeta^2}\right)$$

$$F_{20}^{(3)} = -\frac{3}{\sqrt{2}}\frac{s}{\zeta}\exp\left(-\frac{s^2}{2\zeta^2}\right), \quad F_{30}^{(3)} = -i\sigma\frac{\sqrt{3}}{2}\exp\left(-\frac{s^2}{2\zeta^2}\right),$$

whence the explicit complex-amplitude expression follows:
$$u_{ST}^{(3)} = \frac{1}{6}\left\{\left(\frac{s}{\zeta}\right)^3 - \frac{3}{2}\frac{s}{\zeta} + 3i\sigma\left[\left(\frac{s}{\zeta}\right)^2 - \frac{1}{2}\right]\frac{x}{b}e^{-i\chi} - 3\frac{s}{\zeta}\left(\frac{x^2}{b^2} - \frac{1}{2}\right)e^{-2i\chi}\right.$$

$$\left. -i\sigma\left[\left(\frac{x}{b}\right)^3 - \frac{3}{2}\frac{x}{b}\right]e^{-3i\chi}\right\} u_{ST}^{(0)} \tag{63}$$

(again, the normalizing multiplier warrants that the STOV energy for $u_{ST}^{(3)}$ satisfies Eq. (24)).

Operating in the similar manner, one can obtain the description of STOVs with any integer charge. To this end, we notice that expressions (61) and (63) can be recast as
$$u_{ST}^{(2)} = \frac{1}{2}\cdot\frac{1}{4}\left[H_2\left(\frac{s}{\zeta}\right)H_0\left(\frac{x}{b}\right) + 2i\sigma e^{-i\chi}H_1\left(\frac{s}{\zeta}\right)H_1\left(\frac{x}{b}\right) + \left(i\sigma e^{-i\chi}\right)^2 H_0\left(\frac{s}{\zeta}\right)H_2\left(\frac{x}{b}\right)\right]u_{ST}^{(0)},$$

$$u_{ST}^{(3)} = \frac{1}{6}\cdot\frac{1}{8}\left[H_3\left(\frac{s}{\zeta}\right)H_0\left(\frac{x}{b}\right) + 3i\sigma e^{-i\chi}H_2\left(\frac{s}{\zeta}\right)H_1\left(\frac{x}{b}\right)\right.$$

$$\left.+3\left(i\sigma e^{-i\chi}\right)^2 H_1\left(\frac{s}{\zeta}\right)H_2\left(\frac{x}{b}\right) + \left(i\sigma e^{-i\chi}\right)^3 H_0\left(\frac{s}{\zeta}\right)H_3\left(\frac{x}{b}\right)\right] u_{ST}^{(0)}$$

where $H_m$ denotes the Hermite polynomial (A4). These representations suggest an immediate way of generalization by using the Hermite-polynomial addition theorem,[33] which can be expressed in the form
$$\left(p^2 + q^2\right)^{|l|/2} H_{|l|}\left(\frac{p\xi + q\eta}{\sqrt{p^2 + q^2}}\right) = |l|! \sum_{n=0}^{|l|} \frac{p^{|l|-n} q^n}{n!(|l|-n)!} H_n(\eta) H_{|l|-n}(\xi)$$

for arbitrary integer $l$ and $0 \leq n \leq |l|$. Obviously, accepting $\xi = s/\zeta$, $\eta \equiv \eta(z) = x/b(z)$, $p = 1$ and $q \equiv q(z) = i\sigma e^{-i\chi(z)}$, the final generalization of Eqs. (61), (63) for an arbitrary STOV order $|l|$ can be formulated in a closed form[37,38] as
$$u_{ST}^{(l)}(x, y, z, s) = \frac{1}{2^{|l|}\sqrt{|l|!}}\left(1 - e^{-2i\chi}\right)^{|l|/2} H_{|l|}\left[\frac{1}{\sqrt{1-e^{-2i\chi}}}\left(\frac{s}{\zeta} + i\sigma\frac{x}{b}e^{-i\chi}\right)\right] u_{ST}^{(0)}. \tag{64}$$

One can persuade that for $|l| = 2, 3$, this expression reduces to Eqs. (61) and (63), and the coefficient enables the universal complex-amplitude normalization in the form (24), i.e.
$$W^{(l)} = \frac{1}{8\pi}\int\left|u_{ST}^{(l)}\right|^2 dxdyds = \frac{|A|^2}{8\sqrt{\pi}}\zeta.$$

The expressions (61), (63) and (64) are specially constructed such that the STOV evolution upon propagation along axis $z$ looks symmetric with respect to the waist plane $z = 0$ (like for the simplest STOV (21) illustrated by Fig. 2), and the "perfect" circular OV is realized in the waist plane. One can readily see that at $z = 0$, $\chi = 0$, and

$$\left. u_{\text{ST}}^{(l)} \right|_{z=0} = \left( \frac{s}{\zeta} + i\sigma \frac{x}{b_0} \right)^{|l|} \left. u_{\text{ST}}^{(0)} \right|_{z=0}. \tag{65}$$

The situation of "focusing" multicharged OV can be analyzed similarly keeping the analogy with Eqs. (25) – (27). For example, the STOV with $|l| = 2$, showing the "perfect" circular or elliptic structure at $z = z_c$, is described by the complex amplitude distribution

$$\left[ u_{\text{ST}}^{(2)} \right]_f = \frac{1}{2} \left[ \left( \frac{s}{\zeta_c} \right)^2 + 2i\sigma \frac{s}{\zeta_c} \frac{x}{b} e^{-i\chi} - \left( \frac{x^2}{b^2} - \frac{1}{2} \right) e^{-2i\chi} - \frac{1}{2} \frac{|\zeta_c|^2}{\zeta_c^2} \right] \left[ u_{\text{ST}}^{(0)} \right]_f$$

$$= \frac{1}{8} \left[ H_2\left( \frac{s}{|\zeta_c|} \right) H_0\left( \frac{x}{b} \right) \left( e^{-i\chi_c} \right)^2 + 2i\sigma e^{-i\chi - i\chi_c} H_1\left( \frac{s}{|\zeta_c|} \right) H_1\left( \frac{x}{b} \right) + \left( i\sigma e^{-i\chi} \right)^2 H_0\left( \frac{s}{|\zeta_c|} \right) H_2\left( \frac{x}{b} \right) \right] u_{\text{ST}}^{(0)}$$

(66)

where $\zeta_c$ and $\left[ u_{\text{ST}}^{(0)} \right]_f$ are determined by Eqs. (25) and (27). The normalized closed form for an arbitrary STOV order can be obtained similarly to Eq. (64):

$$\left[ u_{\text{ST}}^{(l)}(x,y,z,s) \right]_f = \frac{1}{2^{|l|} \sqrt{|l|!}} \left( e^{-2i\chi_c} - e^{-2i\chi} \right)^{|l|/2} H_{|l|}\left[ \frac{1}{\sqrt{e^{-2i\chi_c} - e^{-2i\chi}}} \left( \frac{s}{\zeta_c} + i\sigma \frac{x}{b} e^{-i\chi} \right) \right] \left[ u_{\text{ST}}^{(0)} \right]_f. \tag{67}$$

This expression coincides with Eq. (64) if $z_c$ corresponds to the waist plane ($z_c = 0$) where $\zeta_c \equiv \zeta$ is real and $\chi_c = \arg \zeta_c = 0$ (cf. Eqs. (25), (A2)).

Using Eqs. (28), (32), (38) – (40) together with (61), (63), (66), (67), one can easily find that the PC and OAM of the multicharged STOV are determined by the analogs of Eqs. (41), (44), (48):[29]

$$x_p = \frac{l}{2k_0} \frac{b_0}{\zeta} = \frac{l}{2k_0 \gamma}, \quad \frac{\Lambda_y}{W} = \frac{l}{2\omega_0} \frac{\zeta}{b_0} = \frac{l}{2} \frac{\gamma}{\omega_0}, \quad \frac{\Lambda_y^{\text{int}}}{W} = \frac{l}{2\omega_0} \left( \gamma + \frac{1}{\gamma} \right). \tag{68}$$

### VIII. GENERATION OF SPATIO-TEMPORAL OPTICAL VORTICES

There are several prospective approaches to the practical STOV generation discussed in literature.[2,14] Conceptually, the most direct method is based on the superposition of properly prepared and phase-shifted non-vortex pulses, similar to Eqs. (18), (19).[9] In principle, this approach is applicable for obtaining any of the complicated STOV structures described by Eqs. (26), (50), (60), (61), (63) – (67), as well as for their generalizations. But it requires preparing a number of special light pulses with prescribed configurations and their precise alignment, which is practically difficult.

Another group of approaches involves manipulations with the STOV Fourier-spectrum, for example, Eq. (28), as well as with the partial spectra over separate spatial or temporal coordinates like Eq. (32). In particular, for the arbitrarily oriented STOV (50) with symmetric conditions (14), the spectral density is described by Eq. (54). This form of spectral density in the Fourier space $(k_x, k_y, k)$ can be obtained if a Gaussian pulse (e.g., of the form (15)) passes an optical system with transmission function

$$T(k_x, k_y, k) \propto -i\zeta\alpha\Delta k - i\beta k_y b + \sigma k_x b. \tag{69}$$

A sort of such transformation is implemented in the 2D pulse shaper[12,13] (Fig. 7(a), (b)) where the general principle of "structuring light in time" (Ref. 39, part 21) is realized. Namely, the input dispersion element performs "spectrum-to-space" transformation such that different spectral components are spatially separated; ideally, each spectral component enters its own spatial channel. Then, in each channel, the prescribed modulation of the spatial (amplitude and phase) and, if appropriate, polarization distributions is executed, after which the spectral channels are recombined by another dispersion element operating in the inverse mode.

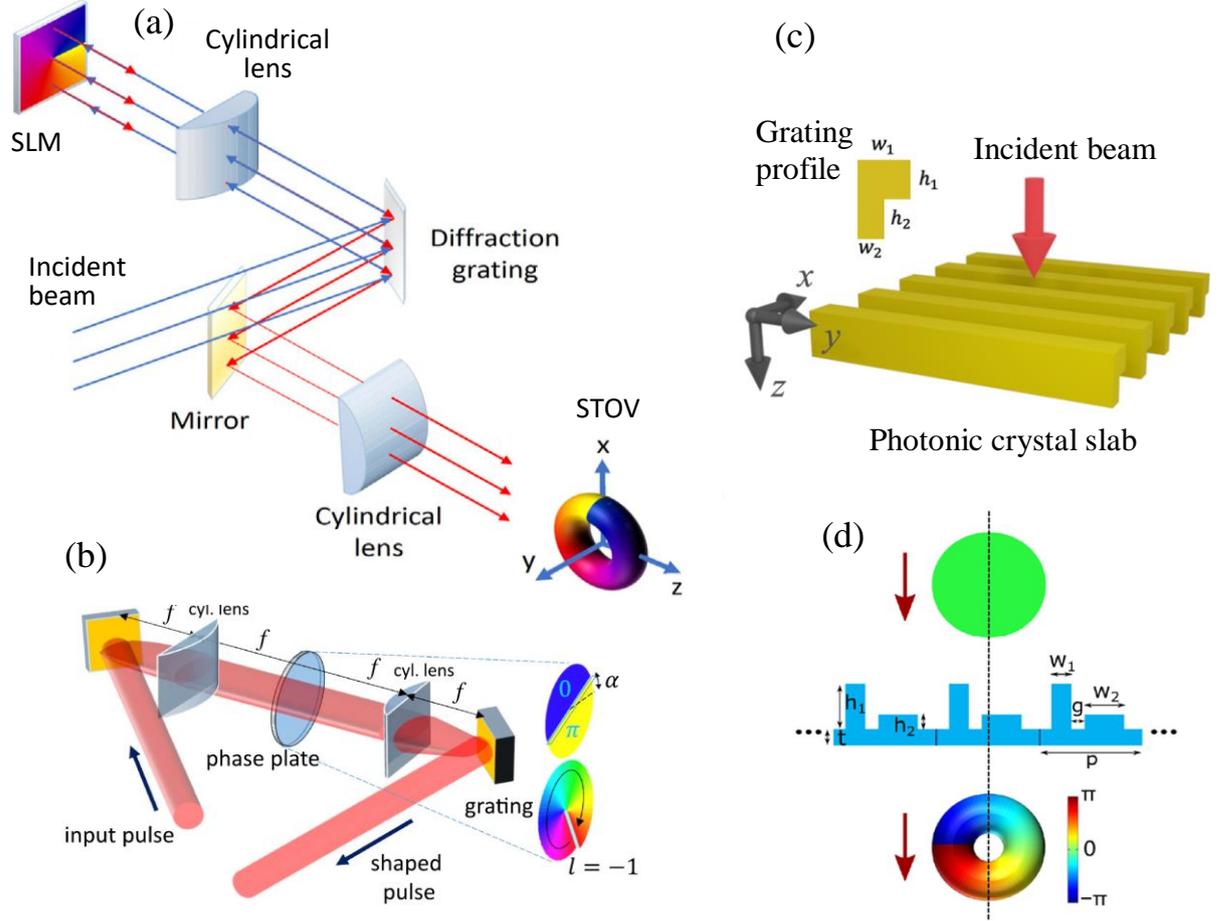

Fig. 7. Examples of the STOV generation principles. (a, b) Beam-shaper schemes with (a) the reflecting SLM [13] and (b) phase plate which perform the spatio-spectral transformation (69) ($\beta = 0$) in the Fourier plane (adapted with permission from Ref. 12 © Optica Publishing Group); (c) photonic-crystal grating with the spectral-depending transmittivity (adapted with permission from Ref. 15 © Optica Publishing Group); (d) ST differentiator with enhanced topological robustness (adapted with permission from Ref. 16 © Wiley).

The schemes of Fig. 7(a), (b) employ the simplified version of this principle. The dispersion elements are diffraction gratings that disperse the input wave packet components with different values of $k$ along the horizontal direction. Then, in the cylindrical-lens focal plane, the complex amplitude distribution is formed proportional to $U^{in}(k_x, k)$ – the Fourier spectrum of the input wave packet $u^{in}(x, s)$. In this plane, the distribution $U^{in}(k_x, k)$ is transformed: in Fig. 7(a), due to reflection at the programmable spatial light modulator (SLM), in Fig. 7(b), upon transmission through the phase plate. The output field recombination is performed by the same (in the reflection scheme Fig. 7(a)) or similar (in Fig. 7(b)) cylindrical lens and grating. In Fig. 7(a), (b), the Fourier-plane modulation introduces the spiral phase distribution which, in vertical direction, is accepted by the spatial $k_x$-components, but its horizontal "part" is imparted to the temporal-frequency components. Accordingly, the transmission $\propto -i\zeta\alpha(k-k_0) + \sigma k_x b$ is approximately realized in the Fourier plane (cf. Eq. (69)), which forms the longitudinal STOV with the transverse OAM at the shaper output. Alternatively, a $\pi$-step phase mask can be placed in the Fourier plane. Depending on the mask orientation angle $\alpha$, the two-lobe structure is realized at the shaper output (see, for example, Fig. 2(a1), (a2) and Fig. 2(e1), (e2)), or Fig. 3(e)), which produces the ring-like or elliptical STOV at a proper cross section of the optical system, e.g. in the far field.[11] This method is rather flexible for generation of STOVs with variable positive or negative topological charges and prescribed ring-like structure localization, dictated by the phase mask or the SLM loading.

But the most universal approaches for the STOV generation involve specially designed metasurfaces.[15] The main element of this arrangement (Fig. 7(c)) is the photonic-crystal slab furnished with the grating formed of material with permittivity $\varepsilon_m = 12$ (the grating profile is shown by the yellow inset). The whole system is polarization-sensitive and is placed between the polarizers with prescribed input and output polarization. With specially adjusted sizes $w_1$, $w_2$, $h_1$, $h_2$ (see Fig. 7(c)), very narrow Fano resonance is excited in the slab due to which its transmission (69) for normally incident light of the central frequency $\omega_0 = ck_0$ vanishes, $T(0,0,k_0) = 0$, but for small deviations it can be expressed via the Taylor expansion

$$T(k_x, k_y, k) \propto a_k(k - k_0) + a_x k_x \tag{70}$$

with, generally, complex coefficients $a_k$ and $a_x$. The phase difference between $a_k$ and $a_x$ is determined by the slab properties and can be adjusted to $\pi/2$, which realizes the transmission function (69) with $\beta = 0$ responsible for the longitudinal OV (21). The terms proportional to $k_y$ appear in the Taylor expansion (70) if the slab is slightly tilted around the $x$- or $y$-axes, and in this manner the STOV with arbitrary orientation (see, e.g., Eq. (50)) can be produced.[15]

Similar approaches are known, employing the unique properties of specially designed nanostructures. For example, a ST differentiator based on 1D periodic silicon structure with two rods per period of different heights and widths has been used[16] to realize the transmission function (70) after which the STOV of the form (21) carrying transverse OAM appears immediately without special Fourier-transforming elements: the necessary transformations happen during free propagation of the pulse. The structure of nonlocal mirror-symmetry-breaking metasurface of Fig. 7(d) is prospective for the STOV topological stability.[14,40] Its properties can be regulated via the rod height $h_1$: It was found that the phase singularity in the transmission spectrum only exists if $h_1$ lies between 238.5 nm and 388.7 nm; in this case, the mirror symmetry is broken and phase singularities appear in pairs. For $h_1$ within this range, the STOV can be generated with the structure stable to the metasurface random deviations and fabrication imperfections.

At last, we cannot omit the huge realm of non-linear optical phenomena which provide multitude of very powerful, flexible and universal means for generation and manipulation of ST light fields, including the STOVs. Because of the limited framework of this presentation, we can only briefly mention some noticeable achievements in this area. In particular, the "predecessors" of STOVs with the transverse OAM were first discovered in 3D "spinning" solitons and trains of "light bullets".[23,41,42] During propagation, such fields show spontaneous focusing and self-focusing effects, associated with the transverse OAM. These phenomena can be treated as the ST analogs and generalizations of the ring-like or knotted singularity lines in monochromatic fields whose local orientation may be orthogonal to the direction of propagation[43,44] (the simplest classical examples are the Airy pattern in the focal plane of uniformly illuminated lens,[45] or the circular transverse phase singularity in the superposition of two coaxial Gaussian beams[5]). Important feature of such versions of STOVs is that they are generated spontaneously, in contrast to other methods discussed in this Section; on the other hand, their control (e.g., obtaining STOVs of a regular structure with the desirable transverse-OAM distribution) remained a difficult task.

Notably, one of the first consistent examples of the STOV concept had been theoretically introduced as a soliton solution of a system of non-linear Schrödinger equations, which describe the light propagation in non-linear waveguides.[10] Likewise, one of the first STOV observations was associated with the non-linear collapse and self-arrest of an intense optical pulse.[11] Further studies have shown the remarkable appropriateness of non-linear approaches to the purposeful STOV generation and transformation, especially in the processes of second-harmonic[46,47] and high-harmonic generation.[48] These phenomena offer exceptional possibilities to transform the STOV-structured light from infra-red to ultra-violet or X-ray range, including special opportunities for transfer of the spin and OAM into ultrashort pulses of femtosecond to attosecond ranges.[2]

## IX. BESSEL-TYPE SPATIO-TEMPORAL OPTICAL VORTICES

Specially prepared field excitations in the ST Fourier-space can be used for generation of the STOV-analogs of the non-spreading Bessel beams.[49,50] For example, if the desirable STOV is localized in the $(x, z)$ plane, the following spectral density distribution should be formed in the $(k_x, k)$ Fourier plane:

$$U(k_x,k) \propto \left[ \frac{k_x b_x + i\sigma\zeta(k-k_0)}{\sqrt{(k_x b_x)^2 + \zeta^2(k-k_0)^2}} \right]^{|l|} \delta\left[ \kappa b_x - \sqrt{(k_x b_x)^2 + \zeta^2(k-k_0)^2} \right] \quad (71)$$

($\kappa$ is the assigned parameter of the STOV shape, $\delta$ denotes the Dirac delta-function). After the Fourier-inversion of (71), we arrive at

$$u(x,s) \propto \int U(k_x, k-k_0) \exp[ik_x x + i(k-k_0)s] \frac{dk_x dk}{(2\pi)^2}$$

$$= \int \delta(\kappa - q) e^{i\sigma|l|\varphi_s} \exp[iq\rho\cos(\varphi_s - \phi_s)] q dq \frac{d\varphi_s}{(2\pi)^2} \quad (72)$$

where the polar coordinates in the $(k_x, k)$ and $(x, s)$ ST planes are introduced according to relations:

$$q = \sqrt{k_x^2 + \frac{\zeta^2}{b_x^2}(k-k_0)^2}, \quad k_x = q\cos\varphi_s, \quad \frac{\zeta}{b_x}(k-k_0) = q\sin\varphi_s;$$

$$\rho = \sqrt{x^2 + \frac{b_x^2}{\zeta^2}s^2}, \quad x = \rho\cos\phi_s, \quad \frac{b_x}{\zeta}s = \rho\sin\phi_s. \quad (73)$$

Performing the integration in Eq. (72) with account for the Bessel function properties,[32] one finally obtains the real-space representation of the Bessel STOV

$$u(x,t) \propto \frac{\kappa}{2\pi b_x} \exp\left[il\left(\phi_s + \frac{\pi}{2}\right)\right] J_l\left(\kappa\sqrt{x^2 + \frac{b_x^2}{\zeta^2}s^2}\right) \quad (74)$$

where $J_l$ denotes the Bessel function of the first kind.[32,33]

So far, we considered the STOV fields in free-space where the dispersion is absent and the wavenumber is just proportional to the frequency, $k = \omega/c$. However, the dispersive conditions are often favorable for generation and support of complicated ST fields. In particular, according to Refs. 49,50, the Bessel STOVs (74) can be realized upon propagation in a dispersive medium with the dispersion law

$$k(\omega) = k_{(0)} + k_{(1)}(\omega - \omega_0) + \frac{1}{2}k_{(2)}(\omega - \omega_0)^2 + \ldots \quad (75)$$

where $k_{(0)} = k(\omega_0)$ is the "central" wavevector (generally differs from $k_0$ in Eqs. (2), (6)), the coefficient $k_{(1)}$ determines the STOV group velocity $v_g = 1/k_{(1)}$, and $k_{(2)} < 0$. Under these conditions, the STOV propagates with the group velocity, and, instead of (1), $s = z - v_g t$. Then, the parameter $\zeta^2/b_x^2$ determining the ST shape of the Bessel-type STOV (74) can be expressed through the dispersion parameters (75) in the form

$$\frac{\zeta^2}{b_x^2} = -v_g^2 k_{(0)} k_{(2)} = -\frac{k_{(0)} k_{(2)}}{k_{(1)}^2}.$$

The STOV form (74) should be supplemented by a proper $y$-dependent part. It is natural to suppose that this $y$-dependent part is described by a certain multiplier $g(y,z)$ depending only on $y$ and $z$, with the separable Fourier image $G(k_y)$ such that

$$U(k_x, k_y, k) = U(k_x, k) G(k_y), \quad u(x, y, s, z) = g(y, z) u(x, s, z).$$

Then, the 3D wave packet of the Bessel STOV can be represented as

$$u(x, y, s, z) \propto \frac{\kappa}{2\pi b_x} g(y, z) e^{il(\phi_s + \pi/2)} J_l\left(\kappa \sqrt{x^2 + \frac{b_x^2}{\zeta^2} s^2}\right). \quad (76)$$

For example, $g(y,z)$ can be taken in the form (13) with omitted $x$-dependent contributions, such that

$$G(k_y) = A\sqrt{2} \exp\left(-\frac{1}{2} b_{y0}^2 k_y^2\right), \quad g(y,z) = \frac{A}{b_y \sqrt{\pi}} \exp\left(-\frac{y^2}{2b_y^2} + ik_{(0)} \frac{y^2}{2R_y} - \frac{i}{2}\chi_y\right). \quad (77)$$

Remarkably, $b_y$, $R_y$ and $\chi_y$ in (77) obey Eqs. (A2) whereas $b_x$ is constant; accordingly, the $(x, s)$-dependent part of this expression does not change upon propagation along the $z$-axis, which justifies the non-spreading character of the Bessel STOV[49] (see Fig. 8(a)).

The "perfect" Bessel STOV of (74), (76) has infinite extent along $x$ and $s$ (or $z$) axes; theoretically, it carries infinite power[51,52] and is practically unrealizable. In real situations, the extent of the beam is limited; keeping in mind the analogy with the spatial Bessel beams,[51,52] one can introduce a ST Gaussian envelope thus forming the Bessel-Gaussian wave packet.[53,54] In the initial plane $z = 0$ (waist plane), such a Bessel-Gaussian STOV can be described by the relation

$$u(x, y, s, z = 0) \propto \frac{\kappa}{2\pi b_x} g(y, 0)$$

$$\times \exp\left[-\frac{1}{w_0^2}\left(x^2 + \frac{b_x^2}{\zeta^2} s^2\right)\right] e^{il(\phi_s + \pi/2)} J_l\left(\kappa \sqrt{x^2 + \frac{b_x^2}{\zeta^2} s^2}\right) \quad (78)$$

where $w_0$ is the "waist radius" of the ST Gaussian envelope. Note that the ST Gaussian envelope also evolves with the wave packet propagation according to the regularities dictated by the waist radius $w_0$ and the central wavenumber $k_{(0)}$ [54] (cf. Eqs. (A2)),

$$w(z) = w_0\sqrt{1 + \frac{z^2}{z_{wR}^2}}, \quad R_w = \frac{z^2 + z_{wR}^2}{z}, \quad \chi_w = \arctan\left(\frac{z}{z_{wR}}\right), \quad z_{wR} = \frac{1}{2}k_{(0)}w_0^2 \qquad (79)$$

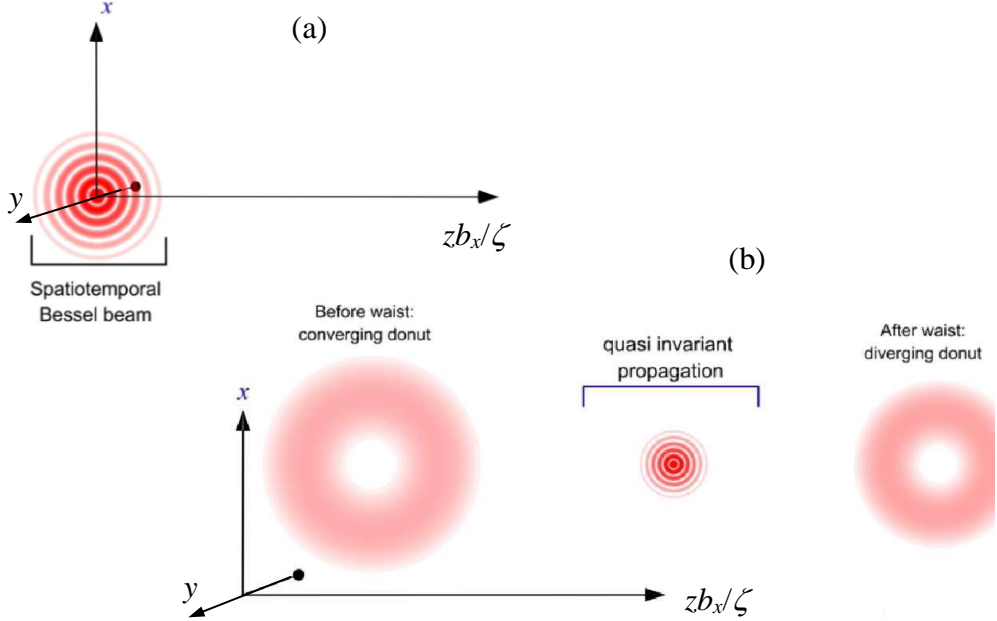

Fig. 8. (Adapted with permission from Ref. 49 © Optica Publishing Group). (a) Propagation of the Bessel STOV (76): the intensity profile (red) in the $(x, s)$ plane moves along the longitudinal axis preserving its initial shape. (b) Propagation of the Bessel-Gaussian STOV (80): the intensity profile in the $(x, s)$ plane evolves under the combined action of the Gaussian envelope transformation (73) and the Bessel profile transformation (81). Near the waist plane, the multi-ring structure is observed whereas a single ring remains far from the waist.

(subscript "$w$" distinguishes the characteristics of the Bessel-STOV Gaussian envelope from the analogous quantities of the Gaussian and Hermite-Gaussian modes, see Eq. (13) and Appendix A). Then, Eq. (78) can be transformed to

$$u(x, y, s, z) \propto \frac{\kappa}{2\pi b_x} g(y, z)$$
$$\times \frac{z_{wR}}{z_{wR} + iz} \exp\left[-\frac{k_{(0)}^2\rho^2 + i\kappa^2 zz_{wR}}{2k_{(0)}(z_{wR} + iz)} + il\left(\phi_s + \frac{\pi}{2}\right)\right] J_l\left(\kappa\rho\frac{z_{wR}}{z_{wR} + iz}\right) \qquad (80)$$

where $\rho$ is determined by Eq. (73) and $f(y, z)$ – by (77). This equation shows that, upon propagation of a Bessel-Gaussian beam, not only the Gaussian envelope width $w(z)$ changes following the standard law (79) but also the radii of the Bessel-beam rings vary according to the relation

$$\rho = \rho_0\left(1 + \frac{z^2}{z_{wR}^2}\right) \qquad (81)$$

where $\rho_0$ is the waist-plane radius. With growing $z$, expression (81) grows faster than the envelope radius $w(z)$ (79), due to which the intensity oscillations localized far from the axis are suppressed (see Fig. 8(b)). Far enough from the waist, the Bessel-Gaussian STOV transforms into a single ST ring in the ($x$, $s$) plane whose radius grows linearly with $z$.[49]

## X. CONCLUDING REMARKS

In this paper, the physical nature, theoretical foundations and experimental realization of the STOV fields are described and briefly characterized. In order to focus on principles and to avoid inessential technical difficulties, the above consideration has been based mainly on the examples associated with paraxial Gaussian (in space and time) wave packets propagating in free space (or linear non-dispersive medium). Additionally, we ignored the fine but meaningful difference between the "space-time formulation" ($z$ denotes a current cross section, $s$ is the time variable specifying the moment when a certain part of the propagating STOV intersects this plane) and "spatial formulation" ($z = ct$ denotes a current moment of time, $s$ is the spatial variable specifying the position occupied by a certain part of the STOV at this moment) of the paraxial ST wave problem.[37,55] The same STOV is described differently in the above formulations but the difference in the corresponding complex amplitude distributions $u(x, y, z, s)$ is of the 1st order of the paraxial-approximation small parameter $\varepsilon$ (see Section II, Eqs. (4) and (7)). For description of the STOV "shape", this difference can be neglected in most cases; however, it is essential for the first-order quantities, such as the transverse momentum components (12). Accordingly, this leads to specific discrepancies in definitions of the "intrinsic" and "extrinsic" OAM,[37,55] which are also omitted in this paper. In Section V, we adhere to the results[17,18,28,29] associated with the "spatial formulation", which look more clear intuitively.

The above restrictions are helpful for understanding basic principles of the STOV structure and evolution but they inevitably put aside a number of other instructive and meaningful examples of the STOV fields. Many of them involve further elaboration and modification of the Bessel STOV model[1,28,29] briefly described in Section IX. Special versions of "ST wave packets"[1] show essential enhancement of the usual STOV properties, combined with the unique propagation-invariant behavior. Such wave packets can be "rigidly" transported in linear media (or in free space) without diffraction or dispersion (preserving the spatial and temporal configuration during the whole evolution); in addition, they can be endowed with controllable group velocities in free space, showing both subluminal and superluminal propagation.

Another important restriction of the presented STOV analysis is the scalar approximation, which, enabling the intuitively clear and meaningful demonstration of the STOV organization and evolution, neglects some fundamental features associated with the vector nature of light waves. Here we can only mention that the vector nature, together with the violation of paraxiality,[56] is very important for some ST topological structures (toroidal and supertoroidal ST pulses, optical skyrmions, hopfions, etc.) which are subjects of recent intense and promising efforts of researchers (Ref. 2, part 11; Refs. 56–59). Such optical structures are essentially singular and topologically determined, and show many exceptional features: non-trivial vector nature with complex orientation of the electric and magnetic vectors; fractal-like and self-similar singular building; essential ST non-separability resulting in non-diffracting propagation over arbitrarily long distances;[60–62] expressive superoscillations[63] (actual field oscillations occur with frequency higher than the highest spectral component); complex and counter-intuitive pattern of the instant energy flows (the regions of anomalous "back flow" may exist where the Poynting vector is directed oppositely to the direction of propagation[64]).

However short and specialized is the present exposition of the STOV properties, it cannot be complete without a brief description on how the specific features of the STOVs can be used in practice. This issue is a subject of discussion so far; however, some impressive steps should be mentioned. First of all, the STOVs can be employed for executing the functions traditionally associated with conventional longitudinal OVs, providing additional benefits of high speed and high energy concentration. The expectable areas of application include optical manipulation,[65] free-space optical communications,[66] space-time differentiators,[16,67] etc. The STOV has been successfully harnessed to manipulate light in nanostructures, for studying the optical properties of the molecular chirality,[68] and supply additional instruments for excitation and investigation of the light-matter interactions.[14] Using the STOVs, optical metrology of nonlinear media, as well as fast processing and transmission of information with intense concentration and release of energy are possible.

An important property of STOVs is the possibility to form the prescribed (ring-like, elliptical or another) structure with the required ST behavior at a given propagation distance. In this regard, the attractive topic for future developments is the ability to control light in different dimensions and degrees of freedom. This is relevant when high-intensity light fields are assigned to control complex ST processes, such as plasma dynamics, dynamics of free electrons and X-ray radiation (Ref. 63, part 2). The problem arises of preparing appropriate radiation sources in the form of a multimode nonlinear laser system that would organize and coordinate the light modes with the desired ST characteristics, and the STOVs can be helpful for its solution.

The non-trivial phase and topological structure of STOVs coupled with the high energy concentration opens interesting prospects for applications associated with the non-linear optical transformations.[63] In particular, in the processes of higher-harmonic generation, a possibility emerges to transform the structured light from infra-red to ultra-violet or X-ray range. A unique opportunity opens up for transferring the spin and orbital AM into ultrashort pulses of femtosecond to attosecond ranges.[63]

Important manifestations of the intrinsic coupling between the spatial and temporal properties of STOVs come to light in the processes of their reflection and refraction at a flat isotropic interface between two media. In this situation, in addition to the conventional Goos-Hänchen and Imbert-Fedorov shifts,[69] a number of new spatial shifts and time delays are found, which are controlled by the value and orientation of the intrinsic AM of the STOV.[70] Due to the special combination of spatial and temporal degrees of freedom in STOVs, time delays and spatial shifts occur even in case of frequency-independent reflection/refraction coefficients, and the "slow" and "fast" propagation of pulses can be realized without the medium dispersion. These results can be important for various problems of scattering of localized vortex states with the help of transverse optical AM, both in classical and quantum formulation.[70]

An interesting version of the STOV, especially suitable due to relative simplicity of its generation, is the partially coherent STOV.[71,72] In contrast to the coherent STOVs, which are obtained using the mode-locking laser pulses, the partially coherent ones originate from the amplified spontaneous emission or from the noise-like pulse states of the fiber laser. In such regimes, the source pulses show some stochastic features that can be modelled by a combination of randomly distributed spectral phase and a Gaussian spectrum profile. The coherence time of such fields exceeds the pulse duration expected from the bandwidth. With growing phase randomness, the regular shape of the partially coherent STOV is destroyed, singularities occur at various ST locations, and multiple amplitude peaks appear in the ($x$, $s$) plane. Parameters of the spatial and temporal coherence of such STOVs are adjustable and can be used for controlling their phase and amplitude structures as well as the singularity position.

Possible applications of the STOVs for data processing are associated with their transverse OAM which adds an additional degree of freedom to the conventional OAM-based information schemes.[73] Moreover, toroidal structures like those described in Fig. 1(d), (e) and Fig. 5, are closely related to particle-like waves such as hopfions,[74] which can be considered as high-dimensional data-carriers with increasing information capacity per pulse for optical communication.[14]

## ACKNOWLEDGMENTS

This work was supported by the Ministry of Education and Science of Ukraine (project # 0122U001830)

## AUTHOR DECLARATIONS

### Conflict of interest

The author has no conflict to disclose.

### Author contributions

**A. Bekshaev**: Writing – original draft (equal); Writing – review & editing (equal).

## DATA AVAILABILITY

The data that support the findings of this study are available from the corresponding author upon reasonable request.

## APPENDIX A: HERMITE-GAUSSIAN MODES

The HG modes (13) and (20), etc., are representatives of the wide family of HG modes[22,24] whose general form is

$$u_{mn}^{AS}(x, y, z) = \frac{1}{\sqrt{2^{m+n}\pi m!n!}} \frac{A}{\sqrt{b_x b_y}} H_m\left(\frac{x}{b_x}\right) H_n\left(\frac{y}{b_y}\right)$$

$$\times \exp\left[-\frac{x^2}{2b_x^2} - \frac{y^2}{2b_y^2} + \frac{ik_0}{2}\left(\frac{x^2}{R_x} + \frac{y^2}{R_y}\right) - i\left(m + \frac{1}{2}\right)\chi_x - i\left(n + \frac{1}{2}\right)\chi_y\right]. \quad (A1)$$

Equation (A1) describes astigmatic (which is indicated by the superscript "*AS*") HG beams with the common waist plane $z = 0$ where the beam size is characterized by the minimal half-widths along the orthogonal axes: $b_x(0) \equiv b_{x0}$, $b_y(0) \equiv b_{y0}$. Normally, these modes are introduced for monochromatic fields; in the current ST context, their expression (A1) is written for the radiation with the central wavenumber $k_0$. Parameters $b_{x,y}$, $R_{x,y}$ and $\chi_{x,y}$ are determined by equations

$$b_j = b_{j0}\sqrt{1 + \frac{z^2}{z_{jR}^2}}, \quad R_j = \frac{z^2 + z_{jR}^2}{z}, \quad \chi_j = \arctan\left(\frac{z}{z_{jR}}\right), \quad z_{jR} = k_0 b_{j0}^2 \quad (j = x, y). \quad (A2)$$

The general asymmetric HG modes (A1), (A2) are used in Eqs. (16). The "classic" HG modes are described by the symmetric versions of the expressions (A1), (A2) where $b_{0x} = b_{0y} = b_0$, which entails the conditions (14):

$$u_{mn}^{HG}(x, y, z) = u_{mn}^{AS}(x, y, z)\Big|_{b_{0x} = b_{0y}}. \quad (A3)$$

These symmetric HG modes in which the beam sizes along the orthogonal transverse directions *x*, *y* are equal, appear in Eqs. (13), (15), (16), (20), (21), (60), (62). The symbol $H_m(\xi)$ denotes the Hermite polynomial of the order *m*; the lowest-order Hermite polynomials are specified by expressions[32]

$$H_0(\xi) = 1; \quad H_1(\xi) = 2\xi; \quad H_2(\xi) = 4\xi^2 - 2; \quad H_3(\xi) = 8\xi^3 - 12\xi. \tag{A4}$$

The normalization constant *A* satisfies the condition

$$\int \left|u_{mn}^{AS}\right|^2 dxdy = \int \left|u_{mn}^{HG}\right|^2 dxdy = |A|^2 \tag{A5}$$

for arbitrary *m*, *n*.

## APPENDIX B: DETAILS OF THE PROBABILITY-CENTER CALCULATION

Here we present some additional data relating the calculations of the PC position. First, let us complete the operations of Section IV and analyze the longitudinal PC coordinate $s_p$. In contrast to expression (39) based on the partial Fourier-transform, now we directly employ the general relation (38):

$$s_p = \frac{1}{W_p} \int \frac{1}{\sqrt{k}} U^*(k_x, k_y, k) i \frac{\partial}{\partial k}\left[\frac{1}{\sqrt{k}} U(k_x, k_y, k)\right] dk_x dk_y dk. \tag{B1}$$

Here the procedure is a bit modified because of the equality

$$\frac{\partial}{\partial k}\left(\frac{1}{\sqrt{k}} U\right) = \left(-\frac{1}{2k^{3/2}} U + \frac{1}{\sqrt{k}} \frac{\partial U}{\partial k}\right) \tag{B2}$$

(arguments of the function $U(k_x, k_y, k)$ are omitted for simplicity of writing). With account for (B2), Eq. (31) leads to the following representation for the integral in Eq. (B1):

$$\int \left(-\frac{i}{2k^2}|U|^2 + \frac{i}{k} U^* \frac{\partial U}{\partial k}\right) dk_x dk_y dk$$

$$= B \int \left\{-\frac{i}{2k^2}\left(\zeta^2 \Delta k^2 + b_0^2 k_x^2\right) + \frac{1}{k}\left[i\zeta^2 \Delta k + \sigma b_0 \zeta k_x - i\zeta^2 \Delta k\left(\zeta^2 \Delta k^2 + b_0^2 k_x^2\right)\right]\right\}$$

$$\times \exp\left[-b_0^2\left(k_x^2 + k_y^2\right) - \zeta^2 \Delta k^2\right] dk_x dk_y dk \tag{B3}$$

where $B = \left(2^{3/2} \pi A \zeta b_0\right)^2$. Note that expression (B1) determines a real quantity, which follows from its definition and is dictated by its form. This means that the imaginary summands present in (B3), inevitably cancel each other, and can be omitted from the beginning, whereas the only real term gives a zero contribution to the integral due to its oddness with respect to the $k_x$ coordinate. Therefore, $s_p = 0$.

Nevertheless, we continue the calculations in order to show that the imaginary terms do vanish eventually. Like in Section IV, the condition (40) applicable to the narrow spectral width is employed. In the resulting expression, that follows from (B3) after substitution of (40), only those summands in figure brackets should be kept which contain even degrees of both variables $k_x$ and $\Delta k$ (the terms with odd degree of at least one variable give zero contributions to the integral). Accordingly, the expression (B3) can be written in the form

$$B \int \left\{-\frac{i}{2k_0^2}\left(\zeta^2 \Delta k^2 + b_0^2 k_x^2\right) - \frac{i}{k_0}\left[\zeta^2 \Delta k^2 - \zeta^4 \Delta k^4 - \zeta^2 b_0^2 k_x^2 \Delta k^2\right]\right\}$$

$$\exp\left[-b_0^2\left(k_x^2 + k_y^2\right) - \zeta^2 \Delta k^2\right] dk_x dk_y dk \tag{B4}$$

(note that all the non-vanishing summands are imaginary). The immediate calculations of the contributions associated with each summand show that they, indeed, mutually cancel, and the whole integral vanishes. This is a good sign; the zero result for (B4) is an argument to support the correctness of calculations. In particular, it confirms the validity of expression (38), in contrast to the formula

$$\begin{pmatrix} s_p \\ x_p \end{pmatrix} = \frac{1}{W_p} \int \frac{1}{k} U^*(k_x, k_y, k) i \begin{pmatrix} \partial/\partial k \\ \partial/\partial k_x \end{pmatrix} U(k_x, k_y, k) dk_x dk_y dk$$

that follows directly from Ref. 29.

Finally, we note that Eqs. (28), (38), (40), (B1) and (B2) allow to derive the expressions for the PC coordinates, containing only the spatial complex amplitude distribution $u(x, y, s)$, without explicit involvement of the spectral densities:

$$x_p = \frac{1}{W_p^{xys}} \int x \left[ |u(x, y, s)|^2 + \frac{i}{k_0} u^*(x, y, s) \frac{\partial}{\partial s} u(x, y, s) \right] dxdyds ; \tag{B5}$$

$$s_p = \frac{1}{W_p^{xys}} \int \left[ \left(s + \frac{i}{2k_0}\right) |u(x, y, s)|^2 + \frac{1}{k_0}\left(is + \frac{1}{k_0}\right) u^*(x, y, s) \frac{\partial}{\partial s} u(x, y, s) \right] dxdyds \tag{B6}$$

$$= \frac{1}{W_p^{xys}} \operatorname{Re} \int s \left[ |u(x, y, s)|^2 + \frac{i}{k_0} u^*(x, y, s) \frac{\partial}{\partial s} u(x, y, s) \right] dxdyds ; \tag{B7}$$

$$W_p^{xys} = \int \left[ |u(x, y, s)|^2 + \frac{i}{k_0} u^*(x, y, s) \frac{\partial}{\partial s} u(x, y, s) \right] dxdyds . \tag{B8}$$

Note that the integral

$$\int u^*(x, y, s) \frac{\partial}{\partial s} u(x, y, s) dxdyds \tag{B9}$$

is an imaginary quantity for physically meaningful functions $u(x, y, s)$, so the expressions (B5), (B8) for $x_p$ and $W_p^{xys}$ are real. Likewise, the line (B6) is also real, despite that it contains some imaginary summands; during calculations, these will mutually cancel, together with the imaginary contributions originating from the term $su^*(x, y, s)(\partial/\partial s) u(x, y, s)$ (as it was demonstrated for the symmetric STOV of Eqs. (21), (31) in expression (B4)). In the line (B7), all such imaginary contributions are formally excluded "from the beginning". For symmetric distributions $u(x, y, s)$ considered in this paper (see, for example, Eqs. (21), (65)), some terms of expressions (B5) – (B8) evidently vanish (e.g., the integral (B9)) but we hope that the "complete" equations (B5) – (B8) would be profitable in applications to more complex non-symmetrical ST wave packets, where the vanishing character of (B9) is not so obvious.

## APPENDIX C: "MICROSCOPIC" FIELD DISTRIBUTION OF THE SPATIO-TEMPORAL OPTICAL VORTICES

The complex amplitude (9), (15), (21), which we operated so far, determines the STOV field distribution averaged over the temporal and spatial (wavelength) oscillation periods. This was sufficient for analysis of the general field profile, its dynamical characteristics, etc., because the oscillation periods are normally rather small and lie beyond the resolution abilities of an observer (cf. Eq. (7)). To describe the genuine "microscopic" field distribution at a certain moment of time, we should return to the initial Eqs. (1) and (2), and explicitly consider the "implied" term $\exp(ik_0 z - i\omega_0 t) = \exp(ik_0 s)$. Then, supposing, for simplicity, $z = 0$, we find

$$E_x(x, y, s) = \operatorname{Re}\left[ u(x, y, s) e^{ik_0 s} \right].$$

The corresponding "microscopic" energy density is proportional to $E_x^2(x, y, s)$;[29] that is, for the STOV (21) under the symmetric conditions (14),

$$E_x^2(x,y,s) = \left(\frac{s^2}{\zeta^2}\cos^2 k_0 s + \frac{x^2}{b_0^2}\sin^2 k_0 s - \sigma\frac{sx}{b_0\zeta}\sin 2k_0 s\right)\exp\left(-\frac{x^2+y^2}{b_0^2} - \frac{s^2}{\zeta^2}\right). \quad (C1)$$

The distribution (C1) differs from that depicted in Fig. 1(a) by the presence of high-frequency modulations ("ripple structure") imposed over the smooth average pattern. Typically for a wave running in the *s*-direction, the ripples represent a series of approximately vertical (*x*-oriented) dark and bright lines alternating along the *s*-axis (two bright lines, corresponding to the wave crest and trough, per one-wavelength distance). However, the regular lines' sequence is distorted by the vortex phase of the STOV. These distortions are poorly visible in "normal" situations, favorable for efficient averaging and considered in the main text (for example, the pattern of Fig. 1(a) contains $\sim 3k_0 b_0/2\pi \approx 500$ wavelengths, and the nearly 1000 bright lines visually merge into a smoothly inhomogeneous spot). That is why in Fig. 9, for better visualization, the case of $b_0 = \zeta = 1.5$ μm is chosen ($k_0 b_0 = k_0\zeta = 15$); similar pictures for Bessel STOVs were presented by K. Bliokh.[29] In this case, the STOV dimensions and the oscillation periods differ not so strongly, and the ripple structure can be explicitly demonstrated. Although such a relation between the field size and the wavelength is marginal for the paraxial approximation and other assumptions of Section II, it supplies a reasonable qualitative picture.

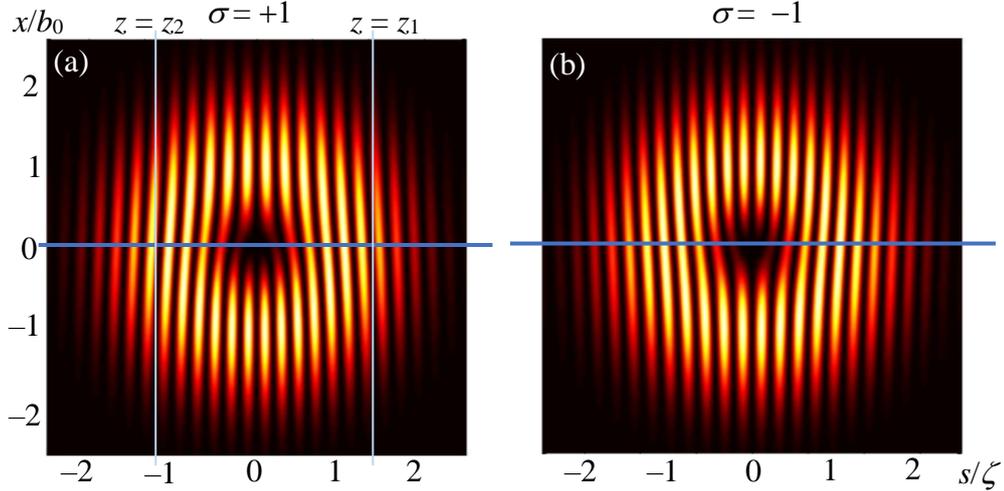

Fig. 9. "Microscopic" energy density distribution (C1) of the symmetric STOV (21) with parameters $b_0 = \zeta = 1.5$ μm, $k_0 = 10^5$ cm$^{-1}$, $z = 0$, for (a) $\sigma = +1$ and (b) $\sigma = +1$. Bright lines correspond to instantaneous crests and troughs of the wave running in horizontal direction; the thin light-blue lines in (a) indicate the positions of fixed transverse plane crossing the leading ($z = z_1$) and rear ($z = z_2$) parts of the STOV packet.

One can see that, e.g., for the positive topological charge ($\sigma = +1$, Fig. 9(a)), an "extra" couple of lines appears at $x < 0$, as compared to the region $x > 0$. This occurs because the vortex phase in the longitudinal plane (*s*, *x*) (see Fig. 1(b)) "supports" the "regular" phase growth $k_0 s$ at $x < 0$, but acts oppositely if $x > 0$. Accordingly, the bright and dark lines are less dense in the upper half-space $x > 0$, which means that the radiation wavelength in the upper half-space is, effectively, larger than in the lower one. This observation naturally explains the spectral-shift effect described by Eq. (33) and Fig. 4(a). Obviously, for the negative STOV (Fig. 9(b)) the situation is opposite and agrees with Fig. 4(b); for higher topological charges $|l| > 1$ (see Section VII), the effect will be enhanced proportionally to *l*.

Additionally, the "microscopic" pictures of Fig. 9 explicitly demonstrate the inevitable asymmetry of a "genuine" STOV structure, even if the averaged energy profile looks symmetric, e.g. as in Fig. 1(a). However, this asymmetry practically does not affect the energy-center position as compared to the value (37) following from the "averaged" definition of Eq. (36). Indeed, for the distribution (C1), the centroid position is determined as

$$\frac{\int \binom{s}{x} E_x^2(x,y,s)\,dxdyds}{\int E_x^2(x,y,s)\,dxdyds} \simeq \begin{pmatrix} 0 \\ -\sigma b_0 \zeta k_0 \exp(-k_0^2 \zeta^2) \end{pmatrix} \simeq \begin{pmatrix} 0 \\ 0 \end{pmatrix}.$$

Here, the $s$-component vanishes due to the symmetry of separate summands of Eq. (C1), the result for the $x$-component is exponentially small (see Eq. (7)) and negligible in the paraxial approximation.

Note also that the bright and dark lines in Fig. 9 can be identified with the local wavefronts. Their inclinations (relative to the "standard" orthogonality to the propagation axis $z$ ($s$)) indicate that the energy flow density (always directed normally to the wavefront[21]) possesses a positive (negative) $x$-component in the right, $s > 0$ (left, $s < 0$) part of the image of Fig. 9(a), which, again, confirms the energy-flow (Poynting-momentum) pattern of Fig. 1(a).

On the other hand, one may recollect another attribute of the directional transverse energy flow in a propagating paraxial field: the "running" behavior of the instantaneous intensity pattern in a fixed transverse plane.[75,76] As an example of such a fixed plane, let us consider the plane $z = z_1$ whose trace in the ($s$, $x$)-plane is denoted by the light-blue vertical line in Fig. 9(a). Obviously, when the STOV propagates towards $z > 0$, the intensity distribution of Fig. 9(a) "moves through" this fixed plane, and simultaneously, due to inclinations of the bright and dark strips, the instantaneous intensity distribution over this plane shows a moving behavior, "running" upward (in the positive $x$-direction). This takes place for any transverse plane crossing the leading side of the STOV, i.e. situated to the right of its center. Oppositely, if the fixed plane crosses the rear side ($z = z_2$ in Fig. 9(a)), the instantaneous intensity pattern in this plane "runs" downward (negative $x$-direction). This picture is in full agreement with the positive and negative $x$-directed energy flows[76] in the STOV illustrated by Fig. 1(a). The similar reasoning, with the proper inversion of signs and directions, is fully correct for the negative-charge STOV (Fig. 9(b)).